%% file: PXoM.tex
\documentclass[conference]{IEEEtran}

\usepackage[available,functional]{ndssbadges}

\input{packages}
\input{macros}

\usepackage{ListStyle}

\title{\huge{Retrofitting XoM for Stripped Binaries without \\ Embedded Data Relocation}}

\begin{document}

\input{authors_2}

\IEEEoverridecommandlockouts
\makeatletter\def\@IEEEpubidpullup{6.5\baselineskip}\makeatother
\IEEEpubid{\parbox{\columnwidth}{
		Network and Distributed System Security (NDSS) Symposium 2025\\
		24-28 February 2025, San Diego, CA, USA\\
		ISBN 979-8-9894372-8-3\\
		https://dx.doi.org/10.14722/ndss.2025.240825\\
		www.ndss-symposium.org
}
\hspace{\columnsep}\makebox[\columnwidth]{}}

\maketitle

\pagestyle{plain}

\input{0_abstract}

\input{1_introduction}

\input{2_background}
\input{3_threat_model}
\input{4_overview}

\input{5_design}

\input{6_security}

\input{7_performance}

\input{8_discussion}

\input{9_conclusion}
\input{ACK_acknowledgement}

% references
{
\balance
%\footnotesize
\bibliographystyle{unsrt}
\bibliography{XOM, randomization, ROP, mpk, benchmark,disassemblers,others, CFI}
}

%\newpage
\input{A_appendix}
\clearpage
\input{A2_artifact_appendix}

\end{document}

%% file: packages.tex
\usepackage[english]{babel}
\usepackage{balance}
\usepackage{graphicx}
\usepackage{tikz}
\usepackage{multirow}
\usepackage{multicol}
\usepackage{booktabs}
\usepackage{url}
\usepackage{cite}
\usepackage{caption}
\usepackage{hyperref}
\usepackage{listings}
\usepackage{mathtools}
\usepackage{amsmath}
\usepackage{xcolor}
\usepackage{soul}
\usepackage[normalem]{ulem}
\usepackage[english]{babel}
\usepackage[flushleft]{threeparttable}
\usepackage[all]{nowidow}
\usepackage{algorithm}
\usepackage{algorithmic}
\usepackage{bbding}
\usepackage{color, colortbl} 
\usepackage{xcolor}
\usepackage{ifsym}

%% file: macros.tex
\renewcommand{\thetable}{\arabic{table}}

\newcommand*\circled[1]{\tikz[baseline=(char.base)]{
            \node[shape=circle,fill,inner sep=1pt] (char) {\textcolor{white}{#1}};}}

\def\RED#1{{\color[RGB]{192,0,0}#1}}
\def\GREEN#1{{\color[RGB]{0,138,63}#1}}

\def\YELLOW#1{{\color[RGB]{201, 145, 4}#1}}
\definecolor{darkgreen}{rgb}{0.0, 0.5, 0.0}

\definecolor{Gray}{gray}{0.9}

\def\CHECKED{{\GREEN{\Checkmark}}}

\useunder{\uline}{\ulined}{}%
\DeclareUrlCommand{\bulurl}{}

\newcommand{\mytool}{{PXoM}}
\newcommand{\MyDisas}{{Unidirectional Disassembly}}
\newcommand{\prevdisas}{{Precise Disassembly}}

\newcommand{\SourceCode}{https://zenodo.org/records/14251050}

\newcommand{\INDSTATE}[1][1]{\STATE\hspace{#1\algorithmicindent}}

%% file: authors_2.tex
\author{

\IEEEauthorblockN{Chenke Luo\IEEEauthorrefmark{2},
Jiang Ming\IEEEauthorrefmark{3},
Mengfei Xie\IEEEauthorrefmark{2}, 
Guojun Peng\IEEEauthorrefmark{2} and
Jianming Fu\IEEEauthorrefmark{2}\IEEEauthorrefmark{1}
\thanks{This paper has been accepted to Network and Distributed System Security (NDSS) Symposium 2025.}
}

\IEEEauthorblockA{\IEEEauthorrefmark{2}Key Laboratory of Aerospace Information Security and Trusted Computing, Ministry of Education,\\
School of Cyber Science and Engineering, Wuhan University}

\IEEEauthorblockA{\IEEEauthorrefmark{3}Department of Computer Science, School of Science and Engineering, Tulane University}

\IEEEauthorblockA{
Email: kernelthread@whu.edu.cn, jming@tulane.edu, \{mfxie96, guojpeng, jmfu\}@whu.edu.cn \\
\IEEEauthorrefmark{1} Jianming Fu is the corresponding author.
}

}

%% file: 0_abstract.tex
\begin{abstract}

System programs are frequently coded in memory-unsafe languages such as C/C++, rendering them susceptible to a variety of memory corruption attacks. 
Among these, just-in-time return-oriented programming (JIT-ROP) stands out as an advanced form of code-reuse attack designed to circumvent code randomization defenses. 
JIT-ROP leverages memory disclosure vulnerabilities to dynamically harvest reusable code gadgets and construct attack payloads in real-time. 
To counteract JIT-ROP threats, researchers have developed multiple execute-only memory (XoM) prototypes to prevent dynamic reading and disassembly of memory pages. 
XoM, akin to the widely deployed W$\oplus$X protection, holds promise in enhancing security. 
However, existing XoM solutions may not be compatible with legacy and commercial off-the-shelf (COTS) programs, or they may require patching the protected binary to separate code and data areas, leading to poor reliability. 
In addition, some XoM methods have to modify the underlying architectural mechanism, compromising compatibility and performance.

In this paper, we present \emph{PXoM}, a practical technique to seamlessly retrofit XoM into stripped binaries
on the x86-64 platform. As handling the mixture of code and data is a well-known challenge for XoM,
most existing  methods require the strict separation of code and data areas via either compile-time transformation or binary patching,
so that the unreadable permission can be safely enforced at the granularity of memory pages.
In contrast to previous approaches, we provide a fine-grained memory permission control mechanism to restrict the read permission of code 
while allowing legitimate data reads within code pages.
This novelty enables PXoM to harden stripped binaries but without resorting to error-prone embedded data relocation.
We leverage Intel's hardware feature, Memory Protection Keys, to offer an efficient fine-grained permission control.
We measure PXoM's performance with both micro- and macro-benchmarks, and it only introduces negligible runtime overhead. Our security evaluation shows that PXoM leaves adversaries with little wiggle room to harvest all of the required gadgets, suggesting PXoM is practical for real-world deployment.

\end{abstract} 

%% file: 1_introduction.tex
\section{Introduction}

The perpetual competition between cyber adversaries and defenders on memory corruption vulnerabilities has intensified,
resulting in an ongoing struggle~\cite{Szekeres13,Kuznetzov14,Hurdle20,PIBE21,ViK22,miTrimmer}.
The prevalence of W$\oplus$X protection (i.e., memory cannot be writable and executable at the same time) in modern operating systems
has led attackers to reuse code snippets from the vulnerable program to construct attacks. Adversaries identify these code snippets, also known as ``gadgets,'' by examining the disassembled binary code~\cite{ROP}. Subsequently, they connect these gadgets in a precise sequence to create harmful payloads and redirect the control flow to the gadgets to launch the attack.
To mitigate this threat, researchers have proposed various code randomization techniques~\cite{Oxymoron,Isomeron,ASLR,AddressObfuscation,bhatkar2005efficient,giuffrida2012enhanced,hiser2012ilr,homescu2013profile,kil2006address,pappas2012smashing,wartell2012binary,williams2016shuffler} to impede the construction of gadgets by reorganizing the code layout in memory. However, code randomization is susceptible to memory disclosure, which makes the randomized code layout evident to attackers and undermines the fundamental memory secrecy assumption of code randomization~\cite{strackx2009breaking}.
The technique of JIT-ROP~\cite{JIT-ROP} leverages repeated exploitation of memory disclosure vulnerabilities to collect code gadgets on-the-fly.
This is accomplished by utilizing the leaked code pointers present on memory pages.
Consequently, JIT-ROP can circumvent code randomization protection, even rendering fine-grained randomization strategies ineffective~\cite{ahmed2020methodologies}.
The premise of JIT-ROP relies on the disclosure of memory pages, where attackers must first traverse disassembled code to gather the required gadgets for the payload construction.
Therefore, a common JIT-ROP defense is to enforce a fine-grained memory permission policy to restrict arbitrary read access to code pages.

Execute-only memory (XoM)~\cite{XnR,LR2,Readactor,uXOM,HideM,NORAX,SECRET} has emerged as a prominent defense against memory disclosure.
By revoking the read privilege of executable memory, XoM deprives attackers of the ability to inspect the code after code randomization has been applied.
XoM implementations have utilized software emulation~\cite{XnR,LR2,uXOM,SECRET} or hardware features~\cite{Readactor,HideM,NORAX} to achieve this objective.
Unfortunately, existing XoM prototypes have failed to gain popularity, and one of the major obstacle comes from the false alarms caused by legal data-in-code reads. 
Ideally, if code and data areas are strictly separated, XoM can safely remove the read privilege only from code sections.
However, for optimization purposes, code-data mixture cases are not rare. For example, compilers may emit data near their accessing code to exploit spatial locality~\cite{CarrSteve94}.

XnR~\cite{XnR} is the first approach to leverage XoM to defend against JIT-ROP attacks, based on the assumption that no data is embedded in the code segment.
Subsequent XoM papers have attempted to address code-data separation in two ways.
The first class of work explicitly separates code and data areas through custom compilers and linkers~\cite{LR2,Readactor,uXOM}.  Obviously, they cannot protect a large number of
legacy and COTS binaries. The second class of XoM work attempts to harden binary code~\cite{HideM,NORAX,SECRET}. Nonetheless, they either rely on debug symbols or error-prone binary patching to
differentiate embedded data from code,  making these approaches impractical.
In particular, HideM~\cite{HideM} modifies the architectural mechanism by segregating all data and code into separate caches. 
This cache mode change has a negative impact on performance and compatibility, as modern CPUs no longer have separate code and data caches. 
These limitations necessitate further research in restricting adversaries' ability to exploit memory disclosure.
On the other hand, Destructive Code Reads (DCR)~\cite{Heisenbyte,NEAR} can tolerate code disclosure by destroying the disclosed code immediately after it is read, thus preventing its execution. 
DCR addresses the challenge of handling legitimate data reads within code pages,
thereby offering enhanced compatibility for protecting binaries.
However, the security guarantees of DCR have been compromised by code inference attacks~\cite{ZombieGadgets}.

\begin{figure*} [t]
    \centering
    \includegraphics[width=1.0\textwidth]{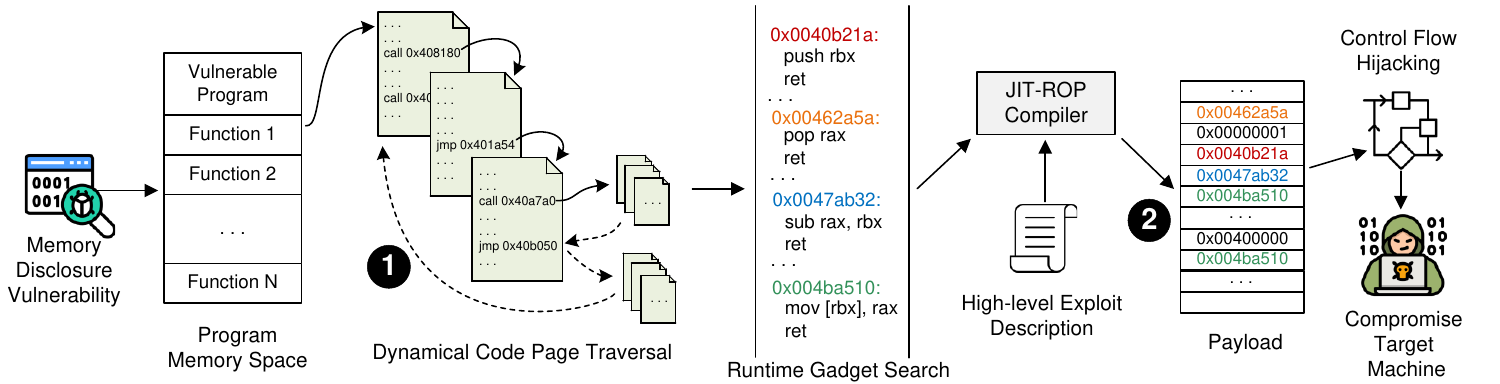}
    \caption{Overview of a typical JIT-ROP attack. Memory disclosure is the premise of a JIT-ROP attack.}
    \label{fig:JIT-ROP}
\end{figure*}

This paper contributes to the ongoing research in XoM policy enforcement by presenting a novel technique called \emph{\mytool}. 
Our approach safeguards stripped binaries from JIT-ROP attacks on the x86-64 platform without the need for embedded data relocation. 
The core of \mytool\ lies in an efficient and fine-grained memory access control policy, which assigns the R$\oplus$X permission to different blocks within a memory page. 
This approach is in contrast to the previous method that required patching of the protected binary~\cite{NORAX}, which involved relocating embedded data out of code pages and updating code-to-data references. 
We note that binary rewriting for relocating embedded data remains a nascent technique, as highlighted in the latest study~\cite{Reassembly23}. 
Our technique enables legitimate data-in-code reads by enforcing the execute-only permission on code areas only, rather than at the granularity of the whole memory page. 
We take advantage of a Intel hardware feature, Memory Protection Keys (MPK)~\cite{IntelManual,libmpk}, to regulate read requests to code areas and embedded data areas at the kernel level, thus minimizing the performance overhead of our approach.

Specifically, to bypass the barrier of precise binary disassembly~\cite{pang2021sok,GroundTruth}, we propose a \emph{\MyDisas} strategy,
which is able to identify all data embedded in code areas without false negatives. 
We customize the binary loader in Linux kernel to load the \mytool-protected binaries, and implement a runtime monitor in
kernel to dynamically scrutinize all read requests to code pages.
To further enhance \mytool's performance on frequently accessed embedded data,
we have developed a cache-like optimization policy.   
Our secure evaluation measures the adversaries' ability to launch a ROP attack. Our results have revealed 
a minimal presence of gadgets in the \mytool-protected binaries, and these leftover gadgets are far from being sufficient to construct a harmful payload.
We conduct a multifaceted performance evaluation with microbenchmarks, marcobenchmarks, and real-world applications,
including lmbench~\cite{lmbench}, SPEC CPU 2006 \& 2017~\cite{SPEC2006,SPEC2017}, three web servers, and four database software.
The results show that \mytool\ only incurs negligible runtime overhead, ranging from  0.22\% to 0.82\% on average.

In a nutshell, we make the following key contributions:
\begin{itemize}
    \item  We propose a new hardware-assisted XoM technique, \mytool, which hardens stripped binaries to impede memory disclosure attempts and eventually
           prevent JIT-ROP attacks. Our work is an advancement in the utilization of hardware features for systems security.
    \item  Our novel fine-grained memory access control policy enable us to overcome the critical limitations of existing work. Our technique allows for
           legitimate data reads in executable memory without necessitating error-prone embedded data relocation.
    \item  To the best of our knowledge, \mytool\ reveals minimal runtime overhead when compared to existing XoM tools. Our extensive evaluation
           demonstrates that \mytool\ is a viable solution for real-world adoption.
\end{itemize}

\noindent \textbf{Open Source }
\mytool's source code and datasets are available at \href{\SourceCode}{\underline{Zenodo}} to facilitate reproduction, replication, and reuse.

%% file: 2_background.tex
\section{Background, Motivation, and Related Work}  \label{sec:background}

In this section, we provide background information on JIT-ROP attacks and the importance of addressing memory disclosure vulnerabilities.
We also review existing approaches for enforcing the XoM policy on userland programs and identify their limitations, which have prompted our research.
Finally, we introduce the hardware feature that we leverage to implement our fine-grained permission control mechanism and kernel-level XoM protection.

\subsection{Overview of Just-In-Time ROP}
With the advancement of fine-grained code randomization~\cite{Oxymoron,Isomeron,giuffrida2012enhanced,kil2006address,williams2016shuffler},
traditional code-reuse attacks~\cite{Roemer12} have evolved into more sophisticated styles like JIT-ROP attacks~\cite{JIT-ROP}, which generate ROP payloads at runtime.
As illustrated in Figure~\ref{fig:JIT-ROP}, a typical JIT-ROP attack consists of two stages.
First, attackers recursively scan code pages using memory disclosure vulnerabilities to search for gadgets (\circled{1} in Figure~\ref{fig:JIT-ROP}),
which typically are code sequences ending with a return instruction.
In the second stage, the collected gadgets are linked together to create a payload that exploits a memory corruption vulnerability (e.g., buffer overflow or use after free)
to hijack the program's control flow (\circled{2} in Figure~\ref{fig:JIT-ROP}).
To complete the search of the whole gadget chain within a small time window (e.g., a few seconds),
JIT-ROP attackers require ``\emph{unfettered access to a large number of the code pages}''~\cite{JIT-ROP} to find usable gadgets quickly.
Therefore, preventing disclosure of memory pages is crucial to mitigating these attacks. 

Programs susceptible to JIT-ROP attacks primarily fall into the following two categories:

\noindent \textbf{Server-side programs},
such as web servers and databases, which allow multiple user interactions, are particularly vulnerable to JIT-ROP attacks. An attacker can interact with the compromised program remotely, incrementally disclosing parts of its code. The search for gadgets and the construction of the malicious payload take place on the attacker's device, while the final payload is executed on the victim's machine.

\noindent \textbf{Client-side programs},
such as Matlab, Autodesk Maya, and JIT engines (e.g., JavaScript), are also vulnerable to JIT-ROP attacks. Attackers exploit these vulnerabilities by utilizing scripting languages. When the victim executes a malicious script, it dynamically searches for code on the victim's machine and constructs the attack payload in real-time. Among these programs, JIT engines are especially prone to exploitation. Attackers can automate exploitation by directing victims to websites hosting malicious scripts, prompting the browser to execute the exploit without user awareness.

\subsection{Execute-only Memory Defense}

The concept of Execute-only Memory (XoM) was once introduced by Multics as early as 1967~\cite{multics_xom}. However, it was not adopted by modern operating systems and hardware until JIT-ROP emerged as an urgent threat.
Next, we introduce XoM approaches designed to protect userspace software from JIT-ROP attacks.

\noindent \textbf{First XoM Defense against JIT-ROP }
The first approach to reintroduce XoM into Linux on the x86 architecture as a defense against JIT-ROP attacks was XnR~\cite{XnR}.
Since there was no hardware feature on x86 supporting XoM, XnR configures the PTE\_PRESENT bit in PTE (Page Table Entry) of code pages as the ``no present'' state, causing all read operations to be intercepted by XnR's page fault handler.
However, due to the substantial overhead incurred by XnR's software implementation, it makes a trade-off to allow several code pages to exist in the present state.
This trade-off causes XnR to miss read operations to these co-existing code pages, leaving memory disclosure opportunities.
More importantly, XnR neglected to handle legitimate read operations that point to code pages. As acknowledged by XnR's authors, XnR will be hindered by such data-in-code reads.
It is clear that the two main factors limiting the deployment of XnR are hardware support and backward compatibility. These challenges have also hindered the broader adoption of XoM since its reintroduction by XnR over a decade ago.

\noindent \textbf{Hardware Support }
When XoM was first reintroduced by XnR, XoM permissions were not supported by hardware and could only be implemented through software emulation, which resulted in significant overhead. 
Later, on the x86 architecture, Intel's Extended Page Tables (EPT)~\cite{IntelManual} hardware virtualization mechanism was utilized to implement execute-only permissions, improving performance to some extent. 
However, EPT requires programs to run within a virtual machine, and virtualization itself introduces additional performance overhead.
Android previously supported XoM on the ARM architecture. However, due to implementation flaws that could lead to the failure of Privileged Access Never (PAN)~\cite{android_pan}, the flawed XoM implementation was removed beginning with Android 11~\cite{android_xom}.
Fortunately, Intel's introduction of Memory Protection Keys (MPK) restored the ability to efficiently separate read and execute permissions, making it the preferred method for implementing execute-only permissions. The Linux kernel has begun supporting execute-only permissions at the kernel level using MPK~\cite{kernel_patch, intel_kernel_xom, csw2023}. In this paper, we also leverage MPK to efficiently enforce execute-only memory permissions. MPK allows us to overcome performance challenges of XoM, enabling us to focus on addressing the other major obstacle to the widespread adoption of XoM: backward compatibility.

\noindent \textbf{Backward Compatibility}
A major factor limiting the widespread adoption of XoM is the challenge of protecting the vast number of precompiled programs. %The primary issue with backward compatibility stems from the mixture of code and data within these programs.
The XnR method~\cite{XnR} is built on the strong assumption that code pages do not contain any data. However, this assumption is not always valid in practice.
Despite modern compilers favoring the separation of code and data, non-code bytes such as jump table data and static read-only data often appear in code sections~\cite{Meng16,Heisenbyte}.
This is confirmed by Pang et al.'s SoK study on mainstream binary disassembly tools~\cite{pang2021sok}, which found that the mixture of code and data is very common in programs.
For instance, the authors discovered 295 hard-coded bytes from the code pages of three test cases and $21,586$ jump tables embedded in the code pages of 57 programs.
Inline assembly code~\cite{Rigger18} in C libraries also frequently embeds data in code sections, such as in the case of OpenSSL, BoringSSL, and FFmpeg, which use handwritten assembly to
speed up their calculations. VirtualBox also employs handwritten assembly to achieve function lazy loading and its virtual extensible firmware interface.
Furthermore, if a binary file links library functions that mix code and data, its code section will also contain embedded data.

The utilization of code and data in conjunction is also required by some security solutions. One such example is KCFI~\cite{KCFI}, which places the hash value of a function's prototype
in the code section via a custom LLVM pass. KCFI reads the embedded hash value to verify the control-flow integrity at runtime.
As admitted by KCFI's developer, it is incompatible with execute-only memory like XnR.

\noindent \textbf{Compile-time Transformation }
In an effort to separate data and code areas for enforcing XoM, one category of follow-up work employs compile-time transformation~\cite{LR2,Readactor,uXOM}.
LR$^{2}$~\cite{LR2} is an example of this approach, which compiles source code using a custom compiler and designates code and data to different memory spaces.
This effectively prevents all read operations to code pages, thereby enabling XoM. However, LR$^{2}$ uses a pure software approach, which involves adding a series of stub code in front of each memory load instruction to verify the legality of read operations, leading to significant overhead. Another solution, Readactor~\cite{Readactor}, also employs a custom compiler and linker to separate code and data.
It utilizes Intel EPT, a hardware-assisted virtualization technique, to manage the read permission of all code pages when mapping the virtual machine's physical address to the host's physical address. 
Finally, uXoM~\cite{uXOM} provides XoM protection on \texttt{ARMv7-M} architecture for embedded devices by manipulating the Memory Protection Unit (MPU).
It does this by implementing a custom LLVM pass 
to convert memory load instructions to unprivileged instructions.

However, all of these solutions require recompiling source code to create code-data-separated binaries, leaving pre-built legacy programs and COTS binaries unprotected.
Furthermore, these approaches cannot cover handwritten assembly functions, which are often used by libraries for enhanced optimization purposes.
For instance, our analysis of OpenSSL 1.1.1q's code section revealed that up to $172,058$ bytes of data are embedded in the code section.

\begin{table*} %[h]
	\caption{Comparison of representative XoM approaches that protect userland programs.}
    \centering
	\label{table:XOMComparison}
	\resizebox{1.0\textwidth}{!}{
        \begin{threeparttable}
		\begin{tabular}{l c c c c c c c c}
			\toprule
			\multirow{2}{*}{ }          & No Source         & No Debug          & Support Data-in-Code      & Hardware\footnotemark[1]      & No Binary             & \multirow{2}{*}{Architecture}     & Runtime               & Memory        \\
                                        & Code Needed?      & Symbols Needed?   & Reads?                    &Feature                        & Patching Needed?      &                                   & Slowdown              & Overhead      \\
		    \midrule
            XnR~\cite{XnR}              & \CHECKED          & \CHECKED          &                           & N/A                           & \CHECKED              & x86/x86-64                        & \RED{8.4\%}          & \GREEN{Negligible} \\
            \midrule
			LR$^{2}$~\cite{LR2}         &                   & N/A               &                           & N/A                           & N/A                   & ARMv8                             & \RED{6.6\%}          & \GREEN{Negligible}    \\
            Readactor~\cite{Readactor}  &                   & N/A               &                           & EPT                           & N/A                   & x86/x86-64                        & \YELLOW{2.8\%}          & \GREEN{Negligible}    \\
            uXoM~\cite{uXOM}            &                   & N/A               &                           & N/A                           & N/A                   & ARMv7-M                           & \RED{7.3\%}          & \RED{High}         \\
            \midrule
            HideM~\cite{HideM}          & \CHECKED          &                   & \CHECKED                  & Split-TLB\footnotemark[2]     & \CHECKED              & x86-64                            & \YELLOW{1.4\%}        & \YELLOW{Small}   \\
            NORAX~\cite{NORAX}          & \CHECKED          &\CHECKED           &                           & AP/XN Bits                    &                       & ARMv8                             & \YELLOW{1.2\%}        & \YELLOW{Small}         \\
            SECRET~\cite{SECRET}        & \CHECKED          &                   &                           & N/A                           &                       & x86                               & \RED{14.4\%}       & \RED{High}       \\
            \midrule
            Heisenbyte~\cite{Heisenbyte} & \CHECKED          & \CHECKED          & \CHECKED                  & EPT                           &                       & x86-64                            & \RED{18.3\%}          & \RED{High}  \\
            NEAR~\cite{NEAR}             & \CHECKED          & \CHECKED          & \CHECKED                  & EPT                           &                       & x86-64/ARMv8                      & \RED{5.7\%}          & \YELLOW{Small}\\
            \midrule
            \mytool                     & \CHECKED          & \CHECKED          &\CHECKED                   & MPK                           & \CHECKED              & x86-64                        & \GREEN{0.36\%}    & \GREEN{Negligible}              \\
			\bottomrule
		\end{tabular}	
        \begin{flushleft}
            \footnotemark[1]{In this column, EPT, TLB, AP/XN, and MPK represents Extended Page Table, Translation Lookup Table, Access Permission, eXecute Never, and Memory Protection Keys, respectively. ``N/A'' means XoM is achieved using page table manipulation~\cite{XnR} or a form of software-fault isolation~\cite{LR2,uXOM,SECRET}.} \\
            \footnotemark[2]{The split TLB technique is not supported anymore by modern x86 processors since the Nehalem microarchitecture (released in 2008).} \\
        \end{flushleft}
    \end{threeparttable}
	    }
\end{table*}

\noindent \textbf{Binary Hardening }
Another category of research aims to enforce the XoM policy with only binary files.
HideM~\cite{HideM} and SECRET~\cite{SECRET} rely on debug information (e.g., function symbols and DWARF) to identify data in code sections prior to runtime.
During runtime, HideM's XoM is achieved by desynchronizing ITLB (Instruction Translation Lookaside Table)
and DTLB (Data Translation Lookaside Table). This causes code and data with the same virtual address to be mapped to distinct physical addresses,
effectively segregating code and data pages. HideM then redirects read operations for code pages to the separate data page. However, this revision disrupts the TLB flush mechanism, leading to performance penalties.
In addition, the split-TLB feature is no longer supported---modern processors released after 2008 have replaced the split-TLB with unified second-level TLBs.
NORAX~\cite{NORAX} disassembles AArch64 stripped binaries and relocates executable data\footnote{NORAX refers to data residing in executable code regions as ``executable data,'' while we refer to executable data as ``embedded data'' in the following sections.} to a non-code segment via binary patching. During relocation, NORAX must correctly update all references to the relocated data. Failing to do so may trigger an access violation and cause protected programs to crash. Unfortunately, updating static data references, such as those from code and the symbol table, is not a simple task~\cite{wartell2011differentiating}. Even more challenging is updating references generated dynamically, such as those from the global offset table (.got) and read-only global data (.data.rel.ro)~\cite{miTrimmer}.
Furthermore, our findings reveal that NORAX's embedded data identification strategy may fail to properly handle cases where code is misidentified as data.
In the event of such an occurrence, NORAX's functionality will cease to operate properly. For instance, if a small function is mistakenly classified as embedded data, the references to the function (e.g., through a function pointer) are also updated to a non-executable area, which may cause the protected program to crash when the function pointer is dereferenced at runtime.

\noindent \textbf{Destructive Code Reads }
To address the issue of XoM methods not supporting legitimate data reads within code pages, 
Heisenbyte~\cite{Heisenbyte} proposed a variant of XoM mechanism, called Destructive Code Reads (DCR). 
Heisenbyte allows memory disclosure but prevents executing the previously disclosed code by destroying the disclosed code right after it is read.
Heisenbyte marks each executable memory page as execute-only and maintains a duplicate copy for each execute-only page. 
When a read operation occurs on the execute-only page, 
Heisenbyte overwrites the read data with random bytes and returns the corresponding data values from the duplicate page. 
Thus, legitimate read operations for data-in-code work correctly, but attackers cannot run disclosed executable memory. 
NEAR~\cite{NEAR} is another DCR approach building upon Heisenbyte, providing a more reliable and efficient memory destruction mechanism.
Although DCR successfully supports legitimate data reads within code pages, 
the code inference attacks proposed by Snow et al.~\cite{ZombieGadgets} have completely undermined DCR's security guarantees.
The core idea of code inference attacks is to disclose a piece of code but not to execute it.
Instead, another piece of code that is strongly related to it will be executed, such as an exact same copy of the disclosed code in a different memory area,
or the relevant code that can be predicted based on the disclosed ones.
Despite the possible evasion to DCR, it still provides valuable insights for advancing XoM. It underscores the critical 
challenge of preventing code exposure while simultaneously permitting legitimate reads to embedded data. 
This inherent dilemma serves as a compelling motivation for our current research.

\noindent \textbf{Comparison of XoM Techniques }
Table~\ref{table:XOMComparison} presents a comparison of various XoM approaches that aim to provide userland software protection.
XnR does not require source code or binary rewriting. However, it doesn't support legitimate reading of embedded data because it assumes no presence of data residing in executable code areas.
Furthermore, the N-page window of XnR leaves an attack surface for adversaries. 
Compile-time transformations require the presence of source code, which fails to protect pre-compiled legacy applications. 
Binary hardening methods, on the other hand, can work with binaries, but only HideM supports the reading of embedded data. 
However, HideM relies on an obsolete hardware feature and changes the normal cache model, making it less compatible. 
NORAX's binary patching may fail to update data references, which changes the original functionality of the protected program. 
DCR methods can work on binaries and support the reading of embedded data, offering the best compatibility among all previous methods. 
Unfortunately, their protections can be bypassed by code inference attacks~\cite{ZombieGadgets}. %rendering them failed to provide reliable protection. 
Additionally, XoM implementations via software emulation, such as LR$^{2}$, uXoM, and SECRET, incur relatively high overhead. 
In conclusion, these limitations highlight the need for further research in developing a practical XoM technique. %for real-world deployment.

In contrast, as demonstrated in $\S$\ref{sec:security} and $\S$\ref{sec:evaluation}, \mytool\ effectively thwarts the disclosure of executable memory while incurring minimal performance and memory overhead.
Besides, \mytool\ does not require source code or debug information.
At last, \mytool\ does not interfere with the original operating system or architectural mechanisms, and unprotected programs remain unaffected by \mytool's kernel components.

\begin{figure*} [t]
    \centering
    \includegraphics[width=0.95\textwidth]{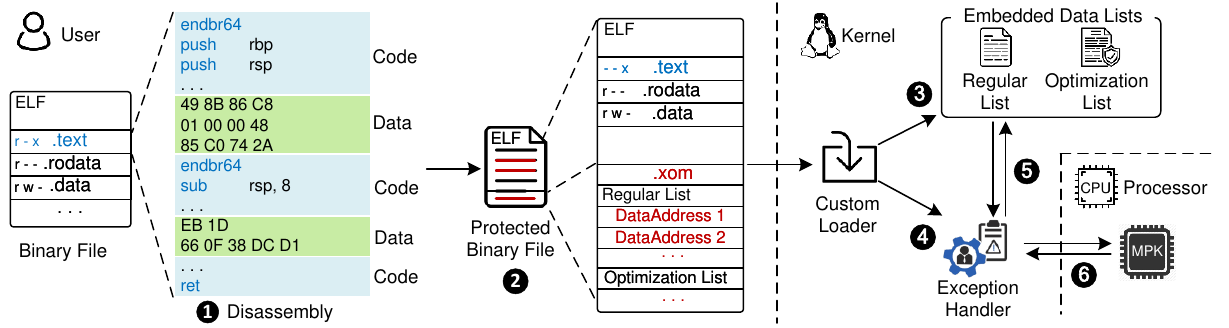}
    \caption{Overview of \mytool.}
    \label{fig:Overview}
\end{figure*}

\subsection{Memory Protection Keys}

Intel Memory Protection Keys (MPK) is a hardware feature that enables stricter permission control on code pages without the need for page table modifications.
The MPK mechanism uses a Protection Key Rights Register (PKRU) to maintain access rights of individual keys associated with specific pages.
It supports three different page permissions: read \& write, read-only, and no access.
Notably, MPK controls read and write permission on memory pages, while traditional permission management mechanisms continue to manage execution permission.
The MPK mechanism can be utilized to configure a memory page's permission as execute-only by disabling the page's read and write permissions.
One of the significant advantages of MPK is its high performance. 
The processors only need to execute a non-privileged instruction (i.e., WRPKRU) to update PKRU, 
which takes less than 20 cycles and does not require any TLB flush or context switching~\cite{SEIMI}. 
However, as MPK keys are localized to each thread, it may lead to inconsistencies between MPK keys of different threads within the same process. 
To ensure the synchronization of execute-only MPK keys between different threads within a process, 
we have utilized the synchronization primitive provided by libmpk~\cite{libmpk}.

\subsection{Kernel-level XoM Protection}
\mytool\ and related works~\cite{XnR,LR2,Readactor,uXOM,HideM,NORAX,SECRET} consider the kernel to be part of
the Trusted Computing Base and thus do not protect against kernel memory disclosure. Another parallel direction
is kernel-level XoM protection, as the kernel itself may be exploited under certain circumstances. For example,
ret2usr attacks~\cite{AModelPopek1978,perfcounter,NvDriverExp,KernelExploit} can redirect control and data flow to user space, compromising the entire system.
KHide~\cite{KHide} and kR\^{}X~\cite{KRX} counteract kernel-level JIT-ROP attacks by enabling XoM protection for kernel memory.
They both rearrange the memory layout of the kernel space, placing executable code in execute-only areas and readable data in read-only areas.
KHide~\cite{KHide} employs the hardware feature Hardware Assisted Paging (HAP) to enforce the XoM policy by mediating access on HAP violation,
while kR\^{}X~\cite{KRX} utilizes Intel Memory Protection Extension (MPX) to enforce XoM permission in a more efficient manner.
However, Intel has discontinued MPX support since the 10th generation of Intel Core processors in 2019~\cite{MPX-dead}.
IskiOS~\cite{IskiOS} simply uses MPK to revoke the read permission of kernel’s code pages.
However, their solution is not applicable to stripped binaries as it does not address the issue of legitimate embedded data reads.

\subsection{Control-Flow Integrity}

A precise implementation of Control-Flow Integrity (CFI) offers significant potential to safeguard applications against ROP attacks by preventing control-flow hijacking.
Currently, hardware mechanisms such as ARM's Pointer Authentication Code (PAC)~\cite{PAC} and Intel's Control-flow Enforcement Technology (CET)~\cite{CET} provide support for CFI enforcement.
Although CFI can still potentially be bypassed under specific circumstances~\cite{OutOfControl,ControlFlowBending,StitchingGadgets,ControlJujutsu,StillDangerous,SizeDoesMatter} and may introduce performance overhead~\cite{CFI_JVM}, it can be deployed alongside other defense mechanisms, thus providing an additional layer of security protection.
From a defense-in-depth~\cite{mughal2018art} standpoint, it is imperative that a critical system incorporates multiple complementary security defenses in practice.

%% file: 3_threat_model.tex
\section{Threat Model}

\mytool\ aims to defend against JIT-ROP by preventing attackers from dynamically disclosing memory, based on a well-defined adversary model.
The model includes the following assumptions:

\begin{itemize}
    \item W$\oplus$X: The target system ensures that the executable and writable permissions cannot coexist on the same memory page. This assumption forms the basis of ROP defenses.
          Otherwise, attackers could simply execute the injected shellcode directly, without the need for ROP techniques.
    \item Randomization: The target program uses a fine-grained code randomization technique, which frustrates adversaries to
         determine the protected program's memory layout in advance.
    \item Control-Flow Hijacking: The target program is vulnerable to memory corruption attacks that allow the adversary to hijack the control flow.
    \item Transparent Configuration: The adversary has knowledge of the target system's configuration, as well as access to the source code of the target program.
\end{itemize}

This adversary model is consistent with previous offensive and defensive papers~\cite{JIT-ROP,XnR,Readactor,HideM}, and specifically aligns with the robust model introduced in JIT-ROP attacks~\cite{JIT-ROP}. 
We exclude side channels and self-modifying binaries protection from our threat model, because they are outside the scope of this paper and are also excluded by other peer works.

Crane et al.~\cite{Readactor} pointed out that there still exists an \emph{indirect memory disclosure attack} that can infer the code layout without directly reading the code pages by harvesting code pointers in stack and heap.
They proposed a method to prevent indirect memory disclosure by redirecting the code pointers to an unreadable trampoline, and thus solved the indirect memory disclosure problem.
As this defense still requires XoM protection to ensure its effectiveness, we focus on addressing the remaining issues in XoM protection.
\mytool\ aligns with the constraint acknowledged by NORAX~\cite{NORAX}, which also works on COTS binaries. 

%% file: 4_overview.tex
\section{Overview}

Our study continues the line of research on retrofitting XoM into stripped binaries.
One of our design goals is to avoid relocating \emph{embedded data} via binary patching. 
To this end, we develop a new fine-grained memory permission control mechanism,
enabling the accommodation of legitimate data-in-code reads.
Figure~\ref{fig:Overview} shows \mytool's architecture that bridges
all layers of the software stack.

\begin{figure*} [t]
    \centering
    \includegraphics[width=0.97\textwidth]{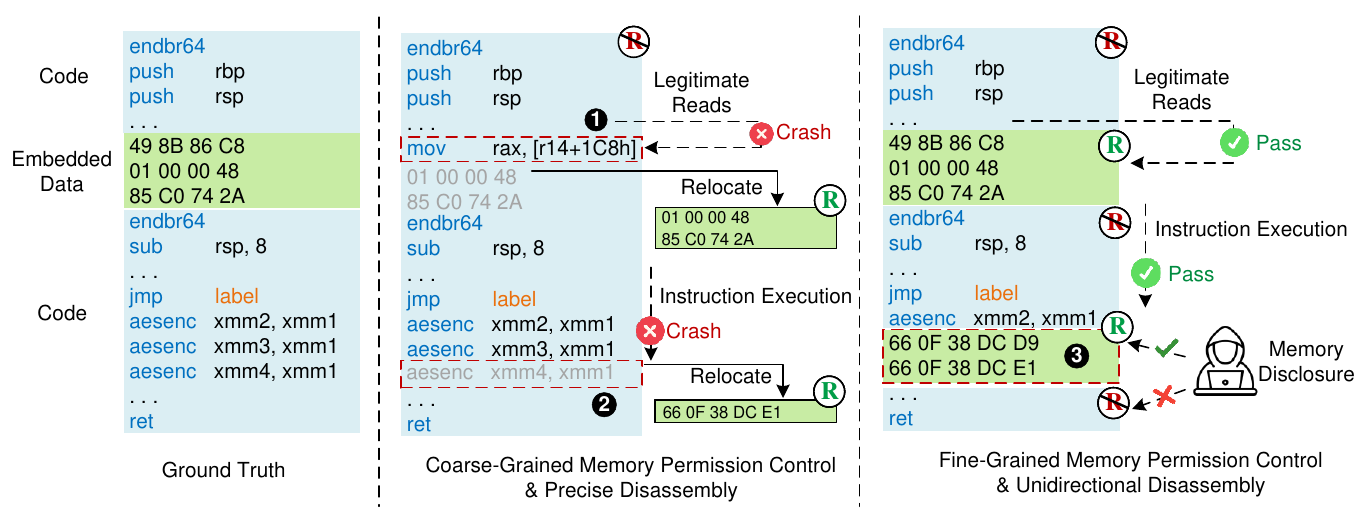}
    \caption{The left side shows a code page containing embedded data.
    The middle section illustrates the previous XoM based on core-grained memory permission control with precise disassembly.
    The right side depicts the effect of PXoM via fine-grained memory permission control with \MyDisas.}
    \label{fig:Disassembly}
\end{figure*}

\vspace*{2pt}
\noindent \textbf{User-space Components }
To determine the areas that are authorized to read, we first identify all embedded data in the binary prior to runtime.
To circumvent the inherent complexity of precisely identifying embedded data, we employ a \MyDisas\ strategy (\circled{1} in Figure~\ref{fig:Overview}).
This disassembly strategy ensures no embedded data will be identified as code.
Subsequently, we append the list of embedded data to the end of the protected binary file and revoke the code segment's read permission (\circled{2}).
In scenarios where data-in-code reads occur frequently, we also customize an optimization policy to speed up the read-legality check.
We create an independent \emph{optimization list} to store the embedded data that are frequently accessed (the right side of \circled{2}).
Please note that in this step, we do not rewrite the binary code. Instead,
we simply add the addresses of embedded data to the end of the binary while marking the code segments as execute-only.

\vspace*{2pt}
\noindent \textbf{Kernel Components }
At the kernel level, we modify the binary loader in Linux kernel to load the protected binary.
In addition to loading the protected binary, the modified binary loader also loads two embedded data lists into kernel-space memory (\circled{3} in Figure~\ref{fig:Overview})
and initializes the exception handler (\circled{4}).
When mapping the code segment, the custom loader loads it as execute-only using the MPK mechanism.
The exception handler is responsible for ensuring the legality of data-in-code reads based on two embedded data lists.
It also dynamically adjusts the optimization list on-the-fly (\circled{5}).
If the address of a read request lies in either the regular list or optimization list, the exception handler will allow this read request.
Otherwise, the read request will be rejected, and \mytool\ will terminate the process and save the context information for further forensics investigation.
The exception handler utilizes the MPK mechanism to efficiently check the legality of read requests (\circled{6}), resulting in very low runtime overhead.

%% file: 5_design.tex
\section{Design} \label{sec:design}

In this section, we follow the workflow of hardening an application to describe each component of \mytool.

\subsection{Fine-Grained Memory Permission Control} \label{sec:identification}

The management of permissions in modern OSs is limited to memory pages, which we call coarse-grained control over memory permissions.  
As a result, previous XoM methods have to relocate embedded data within code pages and update all references to ensure that the program can access them. 
However, the precise identification of code and data within binary remains an undecidable problem~\cite{wartell2011differentiating}.
Previous disassembly efforts~\cite{DEEPDI,ProbabilisticDisassembly,DatalogDsiassembly,andriesse2016depth,bauman2018superset,XDA} aimed to minimize both code-to-data and data-to-code misidentifications, which we can refer to as \emph{\prevdisas}. 
For example, the left side of Figure~\ref{fig:Disassembly} shows a code page containing an embedded data block,
while the middle section displays the \prevdisas\ result of this code page.
In previous XoM methods, erroneous identification of code as data (\circled{2} in Figure~\ref{fig:Disassembly}) leads to the inadvertent relocation of code outside code pages, thereby altering program semantics.
On the other hand, when embedded data are misinterpreted as code, the legitimate read of this data will be prohibited (\circled{1} in Figure~\ref{fig:Disassembly}). 
Both of these errors pose significant crash risks.
Moreover, updating references to relocated embedded code presents a substantial challenge.
Failure to update references to embedded data following relocation may result in program crashes when attempting to read these segments.

We have implemented a fine-grained memory permission control mechanism to assign different permissions to various memory regions within the same memory page. 
This mechanism enables the removal of read permissions for code segments while retaining read permissions specifically for embedded data within the same memory page. 
This approach serves the dual purpose of safeguarding code against disclosure while facilitating legitimate reads of embedded data.
We capture all read requests in executable areas and scrutinize their legitimacy in the kernel's page fault exception handler,  which we will detail in $\S$\ref{sec:handler}.
However, as previously noted,  employing \prevdisas\ may result in both code-to-data and data-to-code misidentifications. 
In the event of code-to-data misidentification, although it may potentially expose small code segments to the risk of memory disclosure, the program can still function correctly because \mytool\ does not relocate embedded data.
Conversely, misinterpreting any data as code may lead to the program crash caused by legitimate read attempts. 
Therefore, we require a disassembly strategy to circumvent the inherent complexity of precisely identifying embedded data, thereby preventing potential crashes.

\begin{figure*} [t]
    \centering
    \includegraphics[width=0.8\textwidth]{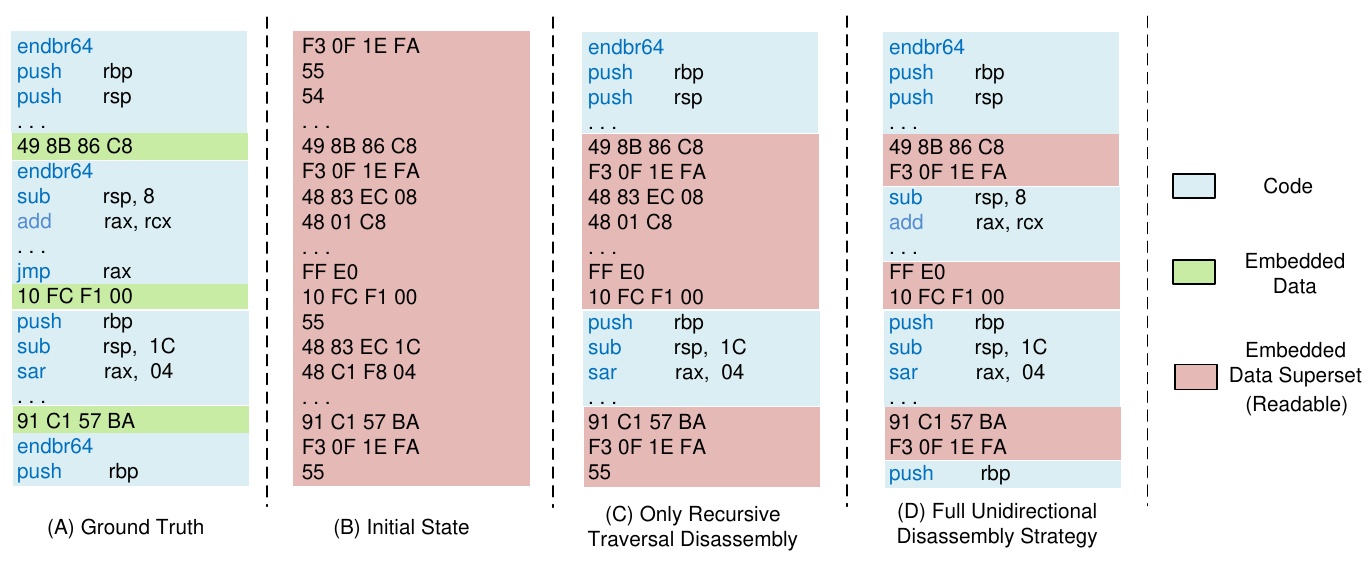}
    \caption{Workflow of \MyDisas strategy. The blue, green, and red sections represent code, embedded data, and the superset of embedded data, respectively. The left side shows a code section containing embedded data. The middle section illustrates the embedded data superset when only recursive traversal disassembly is conducted. The right side demonstrates the minimized embedded data superset after applying the full \MyDisas\ strategy.}
    \label{fig:Unidirectional}
\end{figure*}

\subsection{\MyDisas\ Strategy} \label{sec:disassembly}

We employ a disassembly strategy designed to avoid misidentifying data as code, while tolerating some code being misidentified as data, in order to meet the requirements of our fine-grained memory permission control mechanism. 
Rather than attempting to precisely identify all embedded data, our strategy identifies a superset of embedded data that includes both all actual embedded data and a very small amount of code. We refer to this approach as \MyDisas.
We use Figure~\ref{fig:Unidirectional}(A), a code section containing embedded data, as an example to demonstrate step-by-step process of the \MyDisas.

Initially, the entire code section is marked as embedded data superset (as shwon in Figure~\ref{fig:Unidirectional}(B)).
Then, we apply the recursive traversal algorithm~\cite{schwarz2002disassembly}, following the control flow from the program entry point to identify the code located on the main paths. 
Recursive traversal disassembly partially meets the requirements of our fine-grained memory permission control mechanism by avoiding any data-to-code misinterpretation. This method performs disassembly exclusively on instructions, tracking the program's control flow and thereby preventing the misidentification of data as code.
We then exclude the identified code from the superset, resulting in a smaller superset (Figure~\ref{fig:Unidirectional}(C)).
However, recursive traversal disassembly struggles with handling indirect calls and unreachable functions~\cite{schwarz2002disassembly}, potentially missing up to \textbf{49.35\%} of the code on average~\cite{pang2021sok}. This can lead to a large embedded data superset, thereby exposing too many readable areas to adversaries.
To further reduce the superset, we conduct multiple additional analyses to uncover missed code entry points, subsequently applying recursive traversal disassembly to these identified entry points. During disassembly from each entry point, the identified code is excluded from the superset, thereby minimizing the embedded data superset (Figure~\ref{fig:Unidirectional}(D)).
Specifically, our analyses include examining jump tables, frame unwind information, address-taken functions~\cite{miTrimmer}, and employing function entry identification heuristics~\cite{Egalito} to identify additional code entry points that were not reached by recursive traversal disassembly.
We provide a detailed algorithm in Appendix~\ref{sec:DisasAlgo}.

After obtaining the embedded data superset, we make the entire superset readable and prohibit read permissions for all remaining executable areas using our fine-grained memory permission control mechanism. This ensures that all legitimate embedded data reads are confined to this superset, while any code disclosure attempts outside of this superset are prohibited.
For example, as shown in the right section of Figure~\ref{fig:Disassembly}, 
no embedded data are misinterpreted as code, but two instructions are misidentified as embedded data (\circled{3} in Figure~\ref{fig:Disassembly}). 
As a result, in addition to real embedded data, a small amount of code also retains the read permission. 
For the sake of convenience, we will refer to the embedded data superset simply as embedded data in the following context, as the entire superset will be made readable, and the misidentification of code as data does not affect the executability of the code.
In $\S$\ref{sec:security}, we will measure the coverage of disassembly results and evaluate whether embedded data are exploitable. 
The results demonstrate that our approach provides sufficient protection against memory disclosure without compromising practicality.

\begin{figure} [t]
  \centering
  \includegraphics[width=0.43\textwidth]{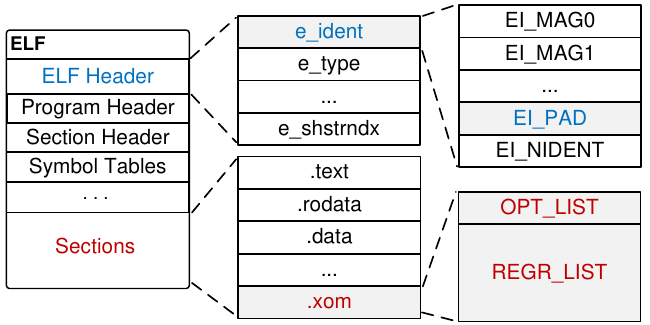}
  \caption{New ELF format for \mytool\ protected binary file.}
  \label{fig:ELFFormat}
\end{figure}

\subsection{New ELF Format}

We define a new ELF format to interact with \mytool's kernel components.
Our minor changes to the original ELF format include utilizing the reserved field and optional section of the ELF file format to store \mytool\ flags and the embedded data list.
Figure~\ref{fig:ELFFormat} shows a visual representation of the new ELF file format we have devised.
First, we use a reserved byte in the ELF header as the \mytool\ specific flag byte, \textbf{XOM\_ENABLE},
to indicate whether the program is protected by \mytool.
This byte is the second byte in the EI\_PAD array, which is a field of e\_ident in the ELF header.
By checking this byte, our custom loader can decide whether to enable \mytool\ protection.
Afterward, the embedded data list is included in an optional section called \emph{.xom}.
The OPT\_LIST and REGR\_LIST represent the optimization list and regular list, respectively.

Please note that the new ELF format remains backward compatible with non-customized loaders because they will disregard the XOM\_ENABLE flag and the .xom section.
The unaltered kernel can execute \mytool-protected binaries without any issues as conventional programs.

\subsection{Custom Loader}

The custom loader is a kernel component of \mytool\ that loads protected binaries and initializes related structures in the kernel.
We provide details about how we organize the kernel structures in Appendix~\ref{sec:KernelStructures}.
To determine if \mytool's protection is enabled, the loader checks the XOM\_ENABLE flag in the ELF header.
If yes, the loader loads the embedded data list stored in the .xom section (\circled{3} in Figure~\ref{fig:Overview})
and initializes an exception handler to ensure data-in-code reads are legitimate (\circled{4} in Figure~\ref{fig:Overview}).
The exception handler is also a kernel component to check the legitimacy of data-in-code reads, and we will introduce it later.
If the \mytool\ flag is not enabled, the standard binary file-loading process takes over.
To prevent attackers from disclosing or tampering with \mytool\ information, all \mytool-related metadata is stored in kernel memory.
The \mytool\ flag, optimization list pointer, and regular list pointer are stored in the \texttt{task\_struct}.
Each process has its own task\_struct object, which stores the context of the process.
Once the lists have been loaded, the custom loader begins mapping the code segment into memory.
During this mapping process, the loader assigns an execute-only PKey, and sets the code section as execute-only with this PKey.
The PKey is a part of the MPK mechanism to set the permission for a group of pages.
This allows our exception handler component to detect any read operation to code pages and determine
if it is a legitimate data-in-code read or a malicious memory disclosure attempt.

\subsection{Exception Handler} \label{sec:handler}

We implement an exception handler based on the original page fault handler for the MPK mechanism to prevent memory disclosure.
With the read permission of all code pages removed via the MPK mechanism, any read request to a code page will trigger a page fault and be caught by our exception handler.
The exception handler then determines if the target address is located in the embedded data areas.
If not, we promptly determine that the running program is under a memory disclosure attack and terminate the compromised process, while saving the context information for further forensics investigation.
Please note that legitimate read operations for data embedded in the code will not trigger \mytool's attack response.
Instead, we take the following actions to allow such a data-in-code read:
1) we restore the read privilege of the target page to allow it to be read temporarily.
2) We set the single-step trap flag to execute only the read instruction and halt at the next instruction.
3) Once the legal read operation is complete, we revoke the code page's read permission again and clear the single-step trap flag to resume the program's normal execution.
To keep track of whether a read operation is legal, we maintain  a flag, called \textbf{XOM\_ALLOW\_READ}, in the \texttt{task\_struct} and set it to false by default.
Once a read operation occurs and is determined to be legal, we set the XOM\_ALLOW\_READ to true.
The single-step trap handler uses this flag to determine whether to allow the read operation.
If true, the data read operation is permitted. 
Subsequently, upon completion of the read operation and restoration of the page permission, the XOM\_ALLOW\_READ  flag is reset to false.

\begin{figure} [t]
   \centering
   \includegraphics[width=0.4\textwidth]{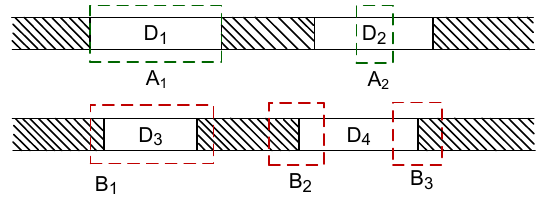}
   \caption{Different types of data-in-code read requests. A read request may access multiple bytes at the same time.}
   \label{fig:ReadRequests}
\end{figure}

Next, we define the legitimacy of data-in-code reads.
A legitimate data-in-code read should not access memory out of the embedded data list.
If a data-in-code read only includes the whole or a subsection of a data area that is in the embedded data list, we take it as a legitimate read.
On the other hand, if a data-in-code read covers a part of memory that is absent from the embedded data list, it is deemed an illegitimate read.
Figure~\ref{fig:ReadRequests} illustrates all legitimate and illegitimate reads.
The $D_1$ $\sim$ $D_4$ represent embedded data areas recorded by the embedded data list, while the shadowed areas represent code areas.
The data-in-code reads $A_1$ and $A_2$ in the first line are legitimate data reads,
while $B_1$, $B_2$, and $B_3$ in the second line are illegitimate data reads.
On the x86-64 architecture, it is possible for a data-in-code read to access multiple bytes using a single instruction.
For instance, the \texttt{MOV} instruction enables the retrieval of up to 8 bytes of data, while Streaming SIMD Extensions allow simultaneous access to a maximum of 16 bytes of data.
$A_1$ covers the entire data block of $D_1$, while $A_2$ covers a portion of $D_2$.
These areas fall within the boundaries of $D_1$ and $D_2$, thereby qualifying as legitimate reads.
Conversely, the regions of $B_1$, $B_2$, and $B_3$ intersect portions of code segments, 
and therefore, they are regarded as memory disclosure attempts.

In the scrutiny of a read request, \mytool\ faces potential performance bottlenecks when iteratively navigating an extensive list of embedded data, a concern exacerbated in programs featuring a large number of code-data interleaving cases.
We built an optimization strategy to address these potential performance bottlenecks, as detailed in the Appendix~\ref{sec:optimization}.

%% file: 6_security.tex
\section{Security Evaluation} \label{sec:security}

In this section, we first examine the outcomes of \MyDisas\ to gauge the completeness of PXoM's protection. 
Following this assessment, we delve into an exploration of PXoM's attack surface to ascertain its effectiveness. 
Through a series of experiments, our findings consistently demonstrate that \mytool\ offers 
a comprehensive defense mechanism against JIT-ROP attacks. %showcasing its ability to significantly bolster the overall security of the system.
Please be aware that the ``embedded data'' in this section is actually the superset of embedded data, as described in \S\ref{sec:disassembly}.

\subsection{\MyDisas\ Result Analysis} \label{sec:correctness}

Our proposed \MyDisas\ strategy ensures the comprehensive coverage of data within code areas, which endeavors to maximize the identification of code areas while ensuring \emph{zero} misinterpretation of embedded data. We present several metrics in this section to gauge the correctness and extent of coverage achieved by this methodology.

The first metric is the Code Coverage, which represents the ratio of disassembled results to the total code bytes.
Code coverage is calculated by Equation~\ref{eq:CodeCoverage} (CC is short for ``Code Coverage.''):
\begin{equation}
	\label{eq:CodeCoverage}
	CC = \frac{Disassembled\ Code\ Bytes}{Real\ Code\ Bytes}
\end{equation}
This metric illuminates \mytool's efficacy in safeguarding the actual code present in binary files. The core of this capability is rooted in PXoM's ability to restrict read access solely to the code sections identified through the disassembly process, while relaxing read access for the remaining regions.
While the real code is more likely to be used as gadgets by attackers, the embedded data could also be used as gadgets under certain conditions.
Hence, we introduce the second metric, the Overall Coverage, which denotes the proportion of disassembled code relative to all executable bytes, comprising real code and embedded data. The Overall Coverage is computed by Equation~\ref{eq:OverallCoverage} (OC is short for ``Overall Coverage.''):
\begin{equation}
	\label{eq:OverallCoverage}
	OC = \frac{Disassembled\ Code\ Bytes}{Real\ Code + Embedded\ Data}
\end{equation}
The third metric is the Number of Embedded Data Blocks, representing how many embedded data blocks in the binaries. The last metric is the Average Embedded Data Block Size. 
These two metrics indicate the distribution density of embedded data blocks within the program.

To gauge the correctness and extent of coverage achieved by our \MyDisas, we use extensive of datasets, including open source applications and COTS closed-source applications, to evaluate the above four metrics. For open source applications, we can extract their ground truth, allowing us to accurately measure these metrics. For the COTS closed-source applications, although we are unable to measure their code coverage (due to the unavailability of their ground truth), we still evaluated their overall coverage, number of embedded data blocks, and average embedded data block size. This is still meaningful in providing supplementary evidence of \mytool's protection capabilities.

\begin{table} [t]
	\centering
	\caption{Disassembly result of \MyDisas\ on open source applications. The ``EDB'' in the fourth and fifth columns represents ``Embedded Data Block.''}
	\label{table:DisassemblyResult}
	\resizebox{0.45\textwidth}{!}{
		\begin{tabular}{l c c c c}
			\toprule
			\multirow{2}{*}{Benchmark}  & Code	        & Overall	    & \#. of       & Avg. EDB  \\   
                                        & Coverage      & Coverage 		& EDB 	       & Size (B)   \\
            \midrule
			SPEC 2017					& 97.58\%       & 96.34\%		& 3020		  & 30 \\
            Webservers                  & 99.39\%		& 98.96\%  		& 593    	  & 9 \\
			Databases                   & 97.67\%		& 98.29\% 		& 9408	  	  & 17 \\
            OpenSSL                  	& 95.61\%		& 86.43\% 		& 8142   	  & 31 \\
			Pang et al.~\cite{GroundTruth} & 96.01\%       & 95.79\%       & 1096     & 52 \\
			\midrule
			Overall 				    & 97.07\%		& 95.29\%		& 4290		  & 31 \\
            \bottomrule
		\end{tabular}	
	    }
\end{table}

\noindent \textbf{Analysis of Results on Open Source Applications }
We evaluate a wide variety of stripped binaries compiled with different optimization options, 
encompassing SPEC CPU 2017 benchmarks, web servers such as Nginx, Apache, and Lighttpd, along with databases including MySQL, MongoDB, Redis, and SQLite. 
We also evaluate a substantial binary dataset obtained from the recent binary disassembly study by Pang et al.~\cite{GroundTruth}.This dataset consists of approximately $4,000$ binary files with a total size of around 20GB, serving as a reliable source of ground truth for disassembly assessments.
Additionally, they released a compilation toolchain capable of extracting ground truth from compiled binaries. We utilized this toolchain to compile the open source applications and extract ground truth.

The evaluation results of above four metrics for open source applications is shown in Table~\ref{table:DisassemblyResult}.
The lowest code coverage, obtained in OpenSSL at 95.61\%, can be attributed to the extensive use of handwritten assembly.
The average code coverage stands at 97.07\%, implying that PXoM can protect a significant portion of the code present within the binary. 
For the overall coverage, OpenSSL still reveals the lowest OC metric at 86.43\%. 
However, the average overall coverage of 95.29\% suggests that PXoM can safeguard the majority of executable memory from potential attackers. 
The fourth and fifth columns of Table~\ref{table:DisassemblyResult} provide insights into the number of embedded data blocks and their average size in bytes. 
The latter indicates the amount of consecutive bytes that attackers can potentially disclose if they manage to identify readable embedded data blocks. 
The average size of an embedded data block is a mere $31$ bytes, which implies that attackers can consecutively disclose only 31 bytes on average when they locate a readable block in the executable area.

\noindent \textbf{Analysis of Results on COTS Closed-Source Applications}
We collected 15 different COTS closed-source applications and analyzed them using \MyDisas. Due to the lack of ground truth information for these closed-source binaries, we are unable to definitively identify the actual code and embedded data, preventing us from reporting a precise Code Coverage metric. However, we assessed other relevant metrics, including Overall Coverage, the number of embedded data blocks, and the average size of these blocks. The results are shown in Table~\ref{table:COTSResults}. The first column of Table~\ref{table:COTSResults} lists the application names and their respective versions used in our evaluation.
The second column presents the Overall Coverage. The lowest Overall Coverage was observed in OracleDB, at 86.44\%. On average, the Overall Coverage is 96.94\%, demonstrating that PXoM can protect approximately 96.94\% of the code in these COTS applications from being exposed to attackers.
The third column displays the number of embedded data blocks, with an average of 1,217 across the analyzed applications. The last column shows the average size of the embedded data blocks, which is 55 bytes. This means that if attackers gain access to an embedded data block with read permissions, they would only be able to read an average of 55 bytes.

\noindent \textbf{Case Studies } Since we cannot obtain the ground truth for COTS closed-source applications, we manually verified some of embedded data and used them as case studies to illustrate how these COTS applications utilize embedded data. For detailed case studies, please refer to Appendix~\ref{sec:COTS_CaseStudy}.

Our experimental results are encouraging, as they validate our claim for embedded data identification. 
These results provide further assurance that PXoM can effectively impose execute-only permission on code areas. 
Next, we will present additional evidence to support that the residual embedded data are insufficient to construct a payload.

\subsection{Attack Surface Analysis}

To tolerate legitimate data-in-code reads, we allow embedded data to remain readable.  
Nonetheless, the disassembly process of \mytool\ may still experience some false positives,
whereby code is misidentified as data. Consequently, the embedded data list provided by \mytool\ includes both the true embedded data and some misidentified code, as shown in the second and third columns of Table~\ref{table:DisassemblyResult}.
Given the embedded data list delivered by \mytool, it is imperative to evaluate adversaries' capabilities to harvest reusable gadgets and subsequently construct an attack payload.
We conduct a gadget search experiment on the embedded data sections for each binary in the dataset utilized in Section~\ref{sec:correctness}. 
The objective of this experiment is to quantify the number of gadgets that can be identified within embedded data regions.

\begin{table} [t]
	\centering
	\caption{Disassembly result of \MyDisas\ on COTS closed-source applications. The ``EDB'' in the fourth and fifth columns represents ``Embedded Data Block.'' The ``VMWare'' represents VMWare Workstation Pro.}
	\label{table:COTSResults}
	\resizebox{0.47\textwidth}{!}{
		\begin{tabular}{l c c c }
			\toprule
			COTS  	                             & Overall	        & \#. of      & Avg. EDB  \\   
            Application                          & Coverage 		   & EDB 	     & Size (B)   \\
            \midrule
            Skype (8.129.0.202)          	     & 99.98\%  		& 718    	  & 58 \\
			DaVinci Resolve (19.0.1)	         & 98.39\%		    & 425		  & 41 \\
            IBM DB2 (15.5.9)            	     & 98.95\%  		& 401    	  & 15 \\
            LiteSpeed (6.3.1)            		 & 99.91\%  		& 43    	  & 68 \\
            Matlab (R2024b)            	         & 98.94\%  		& 690    	  & 125 \\
            AutoDesk Maya (2025\_2)       		 & 98.67\%  		& 1254    	  & 17 \\
            OracleDB (193000)            	     & 86.44\%  		& 577    	  & 31 \\
            Spotify (1.2.45.454)           	     & 99.75\%  		& 1637    	  & 39 \\
			Intel DPC++ (2.1.79)                 & 92.59\% 		    & 2233	  	  & 68 \\
            Intel Fortran (2.1.80)            	 & 92.19\% 		    & 2559   	  & 75 \\
            Steam (1726604483)          		 & 99.27\%  		& 1190    	  & 16 \\
            TeamViewer (15.58.4)           		 & 94.49\%  		& 2461    	  & 25 \\
            Unity (6000.0.20f1)           		 & 98.90\%  		& 434    	  & 51 \\
            VMWare (17.6.0)           		     & 98.65\%  		& 780    	  & 109 \\
            Zoom (6.2.0)          		         & 97.00\%  		& 2740    	  & 90 \\
			\midrule
			Overall 				   	         & 96.94\%		    & 1217		  & 55 \\
            \bottomrule
		\end{tabular}	
	    }
\end{table}

\vspace*{2pt}
\noindent \textbf{ROP Gadget Search }
After applying the ROP gadget search tool ROPGadget~\cite{ROPGadget}, we found that available gadgets are a rare commodity, even for binaries that contain a significant amount of embedded data.
On average, only \emph{seven} gadgets can be extracted from embedded data, which are comprised of $287$ small blocks.
That means these gadgets are distributed among $287$ different locations throughout the entire code section.
These seven gadgets represent the upper limit of potential adversary exploitation. With fine-grained randomization enabled, adversaries lack prior knowledge of where the data blocks are distributed, making it extremely difficult to disclose all the gadgets at once.

As embedded data consist of small data blocks distributed in the code section, 
we present the average embedded data block size in the last column of Table~\ref{table:DisassemblyResult}.
For the overall dataset, the average embedded data block size is only $31$ bytes, and the total of embedded data accounts for 4.71\% (i.e., 1-95.29\%) of the whole code section.
This indicates that if an adversary were to choose an address randomly in the code segment and attempt to disclose code, the probability of this address landing in the readable area is only 4.71\%.
If attackers are fortunate enough to find an embedded data block that can be read through this 4.71\% probability, then the average amount of data they can disclose is only $31$ bytes.
This is insufficient to build a ROP chain, as previous ROP gadget search papers have supported~\cite{schwartz2011q,Microgadgets}.
The ``microgadgets'' technique~\cite{Microgadgets} attempts to use the gadgets restricted to 2 or 3 bytes in length to construct the ROP chain; however, it needs to scan at least 3MB of code to find enough microgadgets.
Schwartz et al.~\cite{schwartz2011q} developed an offline verification technique to facilitate ROP attacks necessitating Turing-completeness,
but it still requires at least 20KB of code to construct a complete payload chain.

\vspace*{2pt}
\noindent \textbf{Case Studies }
To show the effectiveness of \mytool\ protection on real-world threats,
we leverage the JIT-ROP attack framework, jitrop-native~\cite{jirop-native}, to exploit a Nginx arbitrary memory disclosure vulnerability. 
We also conducted an experiment to show that the WRPKRU instruction does not pose a threat to \mytool\ protection.
Please refer to Appendix~\ref{sec:casestudy} for details.

Our comprehensive experimental results demonstrate that \mytool\ can effectively safeguard programs against the threat posed by JIT-ROP attacks.

\begin{figure*} [t]
	\centering
	\includegraphics[width=0.99\textwidth]{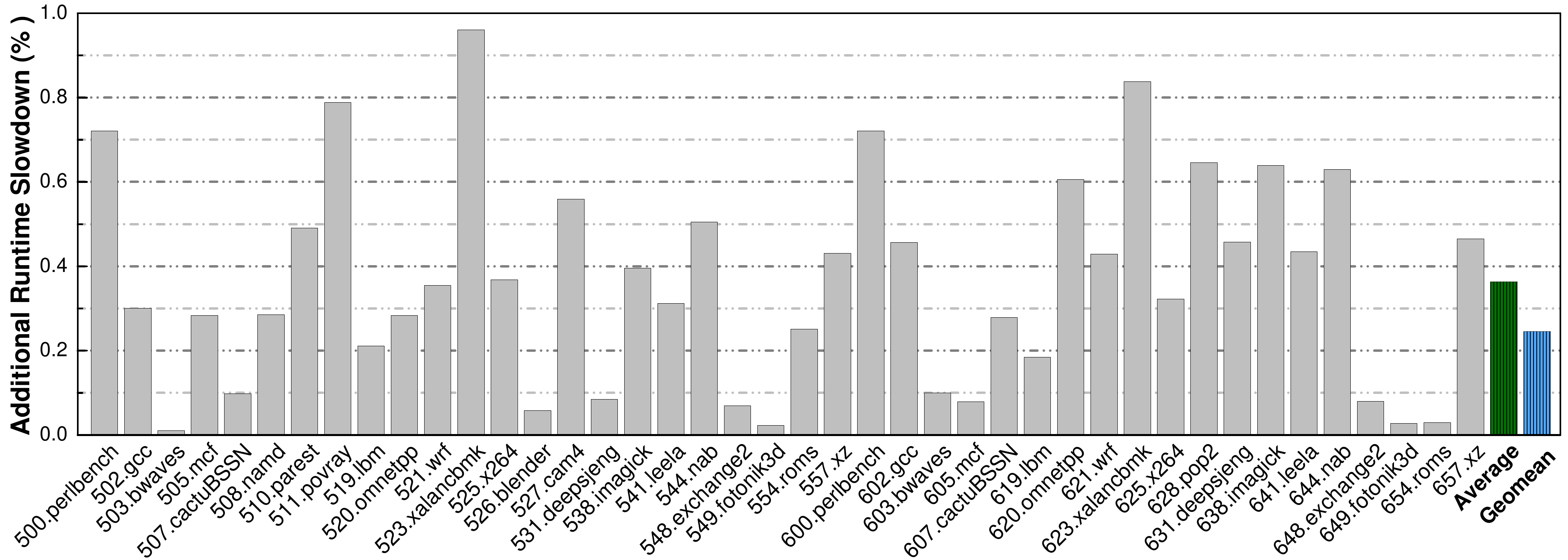}
	\caption{Additional runtime slowdown (\%) of \mytool\ on SPEC CPU 2017 (\emph{ref} workload).}
	\label{fig:SPEC2017}
\end{figure*}

%% file: 7_performance.tex
\section{Performance Evaluation} \label{sec:evaluation}

Our performance experiments evaluate \mytool\ from five aspects: 
1) performance on microbenchmarks;
2) performance on macrobenchmarks;
3) investigation into overhead causation;
4) performance on programs exhibiting high-frequency embedded data reads and the impact of our optimization;
5) performance comparison with existing work.

\begin{table} [t]
	\centering
	\caption{Time for kernel operations related to process creation and page fault handling (in $\mu$s). Smaller is better. 
          In the first row are the names of benchmarks in lmbench. The ``Prot Fault*'' in the last column shows the overhead 
          when data-in-code read-legality check is triggered.}
	\label{table:MicroPXoM}
	\resizebox{0.43\textwidth}{!}{
		\begin{tabular}{c c c c c c}
			\toprule
			\multirow{2}{*}{Kernel}   & Fork    & Exec      & Page     & Prot     & Prot     \\
			                          & Proc    & Proc      & Fault    & Fault    & Fault*   \\
      \midrule
      Standard                  & 109     & 339       & 0.776    & 0.484    & 0.484      \\
			PXoM                      & 110     & 341       & 0.780    & 0.487    & 1.548       \\
			\midrule
      Overhead                  & 0.92\%  & 0.60\%    & 0.52\%   & 0.62\%   & 3.20X       \\
      \bottomrule
		\end{tabular}	
	    }
\end{table}

Our evaluations were conducted on a desktop machine featuring Intel Core i9-13900KF and 64GB of memory, running Ubuntu 23.10 with Linux kernel 6.5.0.
Our evaluation results indicate that \mytool's protection results in negligible additional overhead, ranging from 0.24\% to 0.82\% on average.
Even in OpenSSL, which contains a substantial amount of data-in-code reads, the overhead caused by \mytool\ is only 0.82\%.
The memory overhead incurred by \mytool\ is also minimal, with an average of only 0.13\%.
The following subsections focus on the measurement of runtime slowdown, whereby we ran both the standard version and the \mytool-protected version for each binary, respectively.

\subsection{Microbenchmarks}

Compared to the standard Linux kernel, we made modifications to the kernel's binary loader, page fault exception handler, and process context structure.
Therefore, we run lmbench~\cite{lmbench} on both the standard Linux kernel and the modified Linux kernel to assess the performance overhead induced by our kernel modifications.

Table~\ref{table:MicroPXoM} shows the running time for kernel operations related to process creation and page fault handling.
The \emph{Fork Proc} and \emph{Exec Proc} are process creation operations using \emph{fork} and \emph{exec}.
They resulted in an overhead of 0.92\% and 0.60\%, respectively, due to the additional steps required for loading XoM metadata.
The \emph{Page Fault} shows the overall page fault handling overhead, with a 0.52\% overhead.
The last two \emph{Prot Fault} show the protection fault handling overhead without and with triggering the data-in-code read-legality check.
The first Prot Fault is the overhead on regular prot fault process, without triggering the read-legality check.
However, the handler still needs to check if the \mytool\ is enbaled, resulting a 0.62\% overhead.
In the second \emph{Prot Fault*}, we deliberately trigger the data-in-code read legality checks to evaluate the performance overhead on legality checking process.
Unlike the first four configurations that incur negligible overhead, this configuration incurs a notable overhead with a 3.20 times slowdown.
However, this seemingly unacceptable overhead does not have a significant impact on the overall performance of programs. 
This is attributed to the interleaving of data-in-code reads with numerous other instructions, 
rendering the overall overhead negligible. A more in-depth analysis of this performance impact will be presented in $\S$\ref{sec:causation}.

We need to store some metadata, such as MPK's PKey and embedded data list, in the process's context structure, 
which may cause overhead during context switches in the kernel.
Table~\ref{table:MicroContextSwitch} shows the context switch time for both standard kernel and modified kernel.
In the first row, the upper half displays the number of processes involved in the context switches, while the lower half shows the memory usage of each process.
For instance, ``2p/16k'' represents a context switch between two processes, each of which uses 16 KB of memory.
All entries exhibit overhead values that are clustered around zero, indicating that \mytool\ does not have a significant impact on the performance of kernel context switches.
We present other lmbench results with low correlation to \mytool\ in Appendix~\ref{sec:lmbench_unrelated}.

\begin{table} [t]
	\centering
	\caption{Context switch time (in $\mu$s). Smaller is better. }
	\label{table:MicroContextSwitch}
	\resizebox{0.48\textwidth}{!}{
		\begin{tabular}{c  c c c c c c c }
			\toprule
			\multirow{2}{*}{Kernel}   & 2p   & 2p   & 2p  & 8p   & 8p    & 16p  & 16P       \\ \cmidrule(lr){2-8} 
			                          & 0K   & 16K  & 64K & 16K  & 64K   & 16K  & 64K       \\
      \midrule
      Standard                  & 2.10     & 2.59    & 2.72     & 2.58     & 2.59     & 2.71     & 2.73 \\
			PXoM                      & 2.13     & 2.61    & 2.70     & 2.64     & 2.65     & 2.72     & 2.71  \\
			\midrule
      Overhead                  & 1.43\%   & 0.77\%  & -0.74\%  & -0.93\%  & 2.33\%  & 0.37\%    & -0.73\%   \\
      \bottomrule
		\end{tabular}	
	    }
\end{table}

\subsection{Macrobenchmarks}

We evaluate the performance impact of \mytool\ on compute-intensive programs using SPEC CPU 2017, the latest generation of SPEC CPU benchmarks with larger and more complex workloads than its predecessors.
We compiled both the standard version of SPEC CPU 2017 and the version that was protected by \mytool,
and ran them using the \emph{ref} workload on our test machine.
We take the running time of standard versions as the baseline to measure the additional overhead incurred by \mytool's protection.
Besides, to compare with the performance data of previous XoM approaches, we also conducted a performance evaluation on SPEC CPU 2006 (see Appendix~\ref{sec:perf-compar}).

Figure~\ref{fig:SPEC2017} shows the runtime slowdown caused by \mytool\ on SPEC CPU 2017,
with the green and blue striped bars on the rightmost side showing the average and geometric mean value of overhead, respectively.
From Figure~\ref{fig:SPEC2017}, we can see that eleven overhead values are very close to zero, while
five benchmarks (ID numbers: 523, 623, 511, 500, and 600) reveal relatively high overhead. The peak overhead value happens in \texttt{523.xalancbmk}.
The two \texttt{xalancbmk} benchmarks (523 \& 623) transform XML documents into HTML, text, or other XML document types.
The \texttt{511.povray} is a ray-tracing program, and the two \texttt{perlbench} benchmarks (500 \& 600) are lightweight Perl interpreters.
As all these five benchmarks contain a lot of switch-loop structures,
we conjecture that frequently reading the jump table to call small handler functions contributes to the larger overhead than the remaining benchmarks.
Nonetheless,  the average and geometric mean overhead of tested SPEC benchmarks are 0.36\% and 0.25\%, respectively, indicating that \mytool\ introduces a negligible performance impact on CPU-intensive programs.
In addition to SPEC CPU 2017, we also demonstrate that
\mytool\ introduces minimal runtime overhead to mainstreams web servers and databases (see Appendix~\ref{sec:realworld_apps}).

However, we encountered a major challenge in the current inability to reproduce or replicate previous XoM results, which is an issue that unfortunately plagues the security field.
None of the preceding XoM studies, to the best of our knowledge, have made their tools publicly available.
In Appendix~\ref{sec:perf-compar}, we conduct a separate experiment on SPEC CPU 2006 in order to compare the performance data of \mytool\ with 
that reported by other prominent peer tools in their respective papers~\cite{XnR,LR2,Readactor,HideM,SECRET,Heisenbyte,NEAR}.
In summary, PXoM still exhibits the lowest overhead (0.30\%) among all compared XoM prototypes.

\subsection{Overhead Causation Analysis} \label{sec:causation}

The performance overhead of PXoM is predominantly influenced by two key factors:
1) size of the embedded data list; and 2) frequency of data-in-code reads.
We introduce the term ``read intensity'' to denote the ratio of data-in-code read instructions to the total number of executed instructions. 
Next, we conduct a quantitative examination of these two factors and elucidate the rationale behind the observed negligible overhead incurred by \mytool.

\begin{figure} [t]
  \centering
  \includegraphics[width=0.45\textwidth]{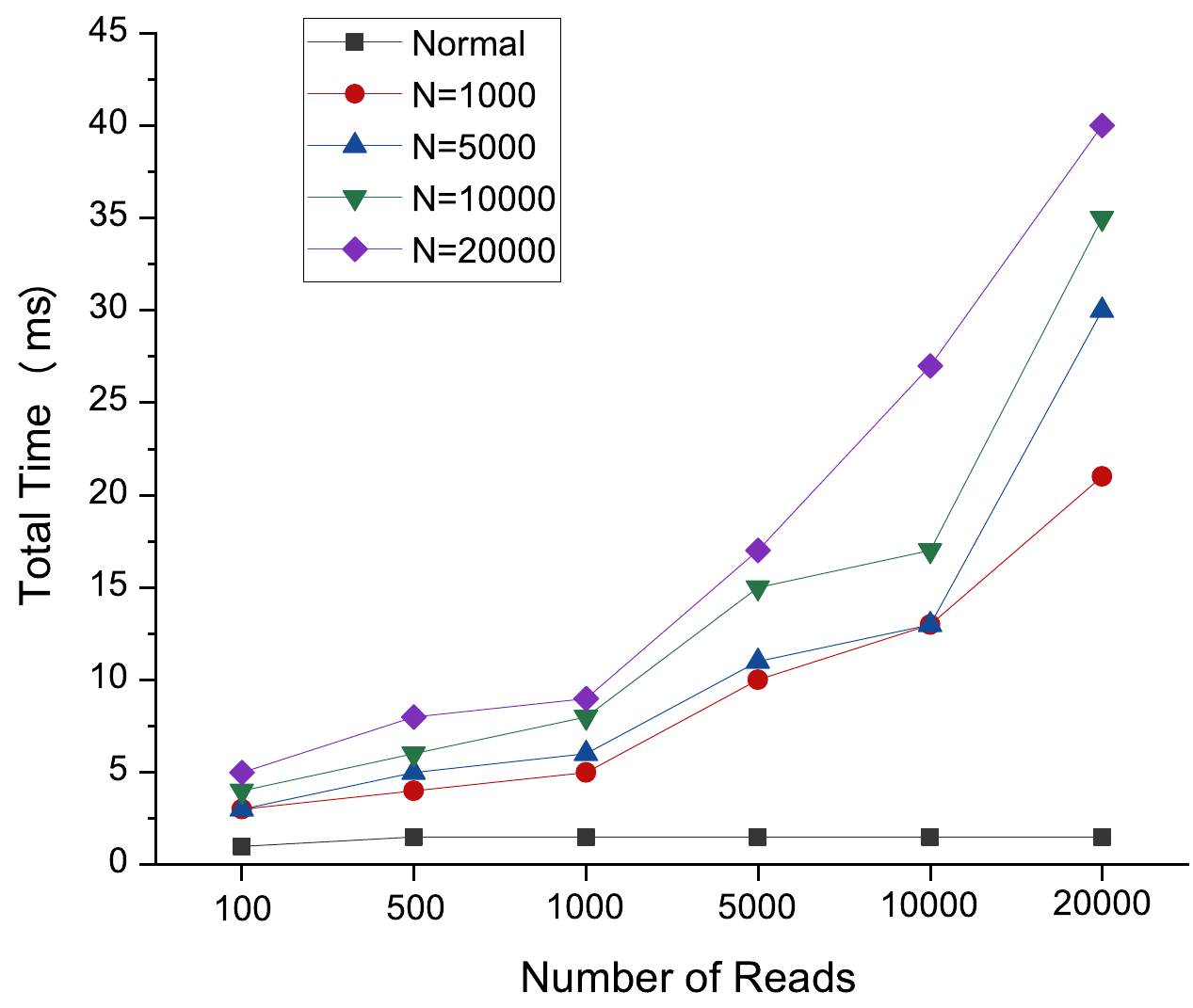}
  \caption{The time for different numbers of embedded data reads for a normal program and PXoM enabled programs with different embedded data list size. 
           The N in the figure means the size of the embedded data list.}
  \label{fig:ReadLatency}
\end{figure}

\noindent \textbf{Read Latency }
Figure~\ref{fig:ReadLatency} illustrates the time taken for performing different numbers of data-in-code reads %$100$, $500$, $1000$, $5000$, $10000$, and $20000$
under different embedded data list sizes (N). 
The horizontal axis denotes the number of data-in-code reads, while the vertical axis represents the time taken to complete the specified number of read requests.
In the ``Normal'' case, which corresponds to disabling PXoM protection, the completion time for read requests remains relatively consistent despite an increase in the amount of reads.
However, in cases where PXoM is activated, an increased volume of read requests correlates with a more pronounced overhead. 
Moreover, a larger size of the embedded data list contributes to a heightened level of overhead.
This observation highlights that when a significant portion of a program's instructions is dedicated to data-in-code reads,
the overhead becomes prominent, especially with larger values of N.
However, in real-world programs, data-in-code read instructions are typically interspersed among a multitude of other instructions. 
Furthermore, as shown in Tables~\ref{table:DisassemblyResult} and Table~\ref{table:COTSResults}, the average size of N (embedded data list size) for both open-source and COTS programs is relatively small, with average values of $4,290$ and $1,217$, respectively.
The consequence is that \mytool\ incurs minimal runtime overhead in real-world programs, as evidenced by the performance data of macrobenchmarks. 
Next, we will quantitatively evaluate the performance impact caused by read intensity.

\begin{figure} [t]
  \centering
  \includegraphics[width=0.36\textwidth]{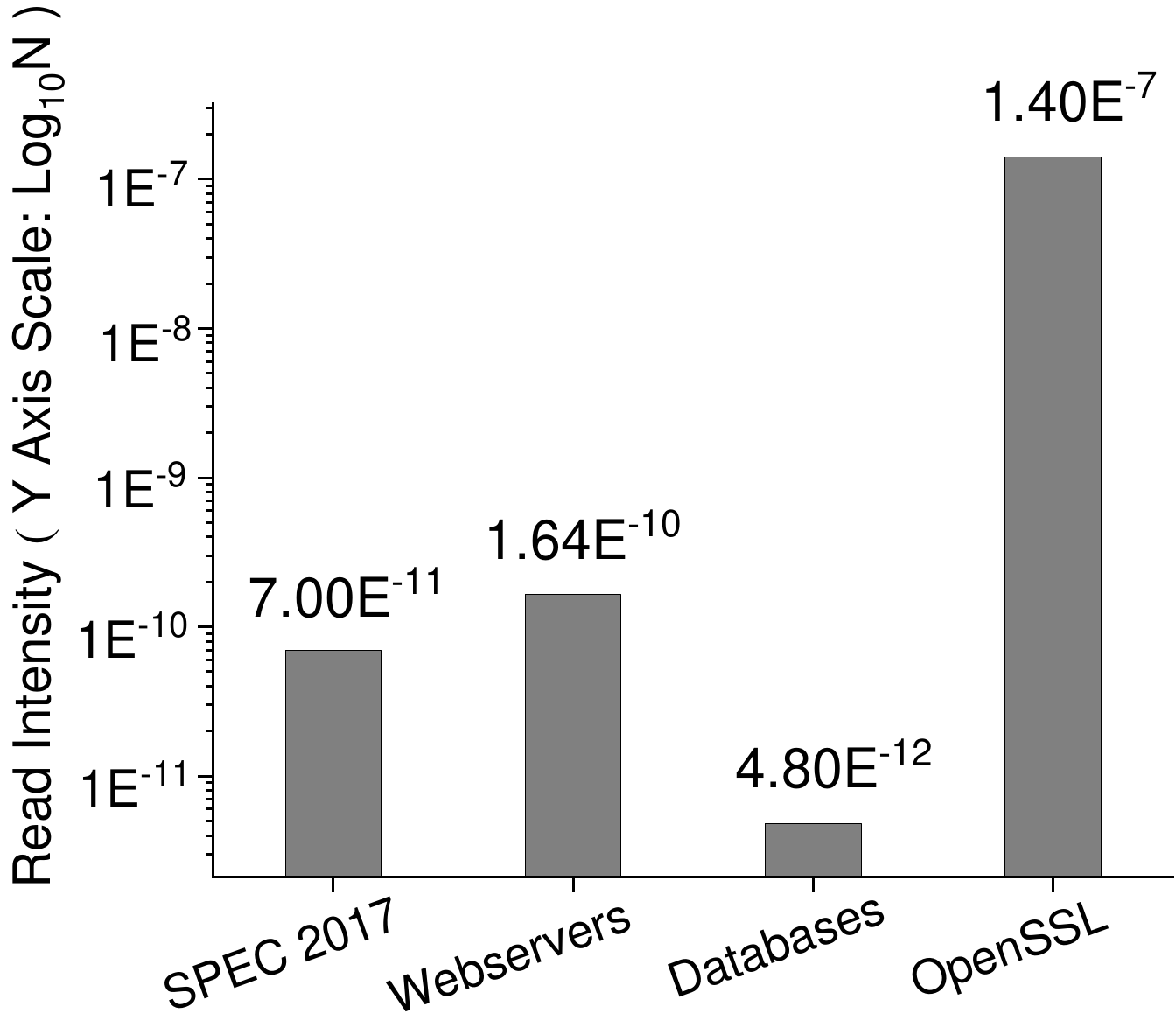}
  \caption{Read intensity of difference benchmarks. The Y-axis values represent the proportion of data-in-code instructions relative to all executed instructions.}  %Y轴数值表示data-in-code指令占所有被执行指令的比例。
  \label{fig:ReadIntensity}
\end{figure}

\vspace*{2pt} 
\noindent \textbf{Read Intensity }
As shown in Figure~\ref{fig:ReadLatency}, there is a direct relationship between a program's intensity of data-in-code reads and the resulting overhead. To capture this, we define the term ``Read Intensity,'' as given by Equation~\ref{eq:ReadIntensity}.
\begin{equation}
	\label{eq:ReadIntensity}
	Read\ Intensity = \frac{\#\ of\ Read\ Requests}{\#\ of\ Executed\ Instructions}
\end{equation}
We have gathered statistics on data-in-code reads and the total number of executed instructions during performance evaluations for SPEC CPU 2017, webservers, databases, and OpenSSL. We have calculated their respective Read Intensity values, as depicted in Figure~\ref{fig:ReadIntensity}.
The program with the lowest Read Intensity is databases, at 4.8E$^{-12}$, indicating that it performs one data-in-code read for every hundred billion instructions on average.
OpenSSL embeds a significant amount of data within the code region to enhance the performance of cryptography algorithms.
It exhibits the highest Read Intensity at 1.4E$^{-7}$. 
Nevertheless, even in OpenSSL, on average, it takes a million instructions to perform one data-in-code read.
The average Read Intensity for all programs is 3.51E$^{-8}$, signifying one data-in-code read is performed after executing ten million instructions on average. This illustrates that while data-in-code reads are not rare in practice, their occurrence rate is extremely low, resulting in \mytool's practical overhead being negligible.

%% file: 8_discussion.tex
\section{Discussion \& Conclusion}

\vspace*{2pt}
\noindent \textbf{Kernel Component Security}
Adding code to the trusted computing base is risky, so we need to pay extra attention to ensuring the security of any additions made to the kernel.
The first potential threat is the user-controllable ``.xom'' section. When the kernel loads a binary, it parses the contents of the ``.xom'' section into the kernel's structures. Improper checks during this process could allow an attacker to exploit the kernel. Therefore, we must conduct careful and strict checks when parsing this list to prevent buffer overflow attacks.
In addition, we added some pointers in the kernel to store data related to XoM. When using these pointers, strict checks must also be enforced to prevent vulnerabilities such as use-after-free and double-free.
Another potential threat is the race condition between different threads. When embedded data reading is allowed, the target code page will be in a readable state for a very short time. If control is taken over by another thread at this time, that thread may disclose the readable memory page. Therefore, it should be ensured that the permission for the reading operation is atomic, and control cannot be taken away during this operation.

\vspace*{2pt}
\noindent \textbf{Protection on Dynamically Loaded Code }
For now, \mytool\ is not designed to protect dynamically loaded (dlopen) code and dynamically generated code, such as in a program running JavaScript code in a JIT engine.
Achieving protection for them will be our future work.
For the dynamically loaded code, the only difference is the loading process.
We will hook the GNU C Library to protect dynamically loaded code.
For the dynamically generated code, when the JIT engine generates JIT compiled code, we can know the location of all embedded data, so we can mark these locations as readable in the JIT engine and apply PXoM protection to the generated code.

\vspace*{2pt}
\noindent \textbf{MPK Security }
While Memory Protection Keys (MPK) provides an efficient mechanism for managing memory permissions, it also raises concerns regarding its own security pitfalls~\cite{Pitfalls}.
Fortunately, new defensive strategies have emerged to further strengthen the security of MPK~\cite{Jenny,Cerberus}.
Ongoing improvements and refinements in this evolving domain continue to enhance the security of MPK.
Many existing works have utilized MPK for sensitive data isolation. For instance, ERIM~\cite{ERIM} and Hodor~\cite{Hodor} utilize MPK to
isolate sensitive data and only allow trusted code to access it by controlling read permissions to these sensitive data areas. 
Similarly, Burow et al.\cite{ShadowStacksMPK} investigate using MPK to provide stronger guarantees for shadow stacks, which are used to make sensitive data on the stack tamper-resistant\cite{ShadowStacks}. 
Additionally, Jin et al.~\cite{CryptoMPK} employ the MPK mechanism to safeguard sensitive key-related data in cryptographic algorithm implementations.

%% file: 9_conclusion.tex
\noindent \textbf{Conclusion}
Execute-only memory (XoM) is a promising solution to prevent memory disclosure and counter JIT-ROP attacks. This paper presents \mytool, a technique that retrofits XoM into stripped binaries without embedded data relocation. Unlike existing approaches, \mytool enables fine-grained memory permission control within a memory page without requiring compile-time transformations or binary patching. Performance evaluations on large programs show negligible runtime overhead, and security assessments suggest \mytool is viable for real-world adoption, potentially shifting the memory defense landscape in favor of defenders.

%% file: ACK_acknowledgement.tex
\section*{Acknowledgment}

We thank our shepherd and all the anonymous reviewers for their valuable comments to improve this paper.
This work is supported by the National Nature Science Foundation of China under Grant No. 62272351, 61972297, and 62172308.
Jiang Ming was supported by NSF grants 2312185 \& 2417055 and Google Research Scholar Award.

%% file: A_appendix.tex
\appendix

\setcounter{table}{0}
\setcounter{figure}{0}
\renewcommand{\thetable}{A\arabic{table}}
\renewcommand{\thefigure}{A\arabic{figure}}

\subsection{Kernel Structure Modifications} \label{sec:KernelStructures}

\begin{figure} [ht]
\begin{lstlisting}[style=CStyle2]
  typedef struct edata{
    struct rb_node node; //The red-black tree node
    uint64_t start; //Executable data block start address
    uint64_t end; //Executable data block end address
    uint64_t freq; // How many time this block was read
  };

  struct pxom_info_t{
    int enabled; //PXoM enable flag
    int read_allowed; //Read operation legality flag
    int pkey; //PKey (in MPK mechanism) of code pages

    struct rb_root regular_list; //The regular list
    struct rb_root optimization_list; //The optimization list
  };
\end{lstlisting}
\caption{The \mytool\ structures in Linux Kernel.}
\label{code:PXoMStructures}
\end{figure}

\begin{figure} [ht]
\begin{lstlisting}[style=CStyle2]
  struct task_struct {
    volatile long         state
    void                  *stack;
    refcount_t            usage;
    unsigned int          flags;
    unsigned int          ptrace;
    (*@ \dots @*)
    struct pxom_info_t   (*@ \RED{pxom\_info} @*);
    (*@ \dots @*)
  };
\end{lstlisting}
\caption{The task\_struct structure in Linux Kernel.}
\label{code:TaskStruct}
\end{figure}

In the Linux kernel, we have defined structures to store essential information about \mytool.
Figure~\ref{code:PXoMStructures} displays these structures, which include fields that correspond to the newly defined ELF format and others that store runtime information for \mytool\ protection.
One such structure is \emph{edata}, which stores the virtual address range of an executable data block in memory.
Another structure, \emph{pxom\_info\_t}, holds basic information about a process protected by \mytool, such as the enable flag, the read permission flag, the PKey assigned to enforce execute-only memory, and the executable data list.
We store the pxom\_info\_t structure in the \emph{task\_struct} structure to save \mytool\ information in the current process context.
Figure~\ref{code:TaskStruct} depicts the task\_struct structure, where we have added the pxom\_info\_t member at the end of the randomized struct fields to ensure compatibility with randomization protection.

\begin{algorithm}
    \renewcommand{\algorithmicrequire}{\textbf{Input:}}
    \renewcommand{\algorithmicensure}{\textbf{Output:}}
    \caption{Unidirectional Disassembly Strategy} 
    \label{alg:unidirectional} 
    \begin{algorithmic}[1]
    \REQUIRE $BC - stripped\ binary\ code$  \\
    $\quad \ \, EP - program\ entry\ point$ 
    \ENSURE $EmbeddedDataSuperset, Abbreviated\ as\ S$
    
    \STATE $S \gets BC$ 
    \STATE $IdentifiedCode \gets RecursiveDisassemble(S, EP)$
    \STATE $S \gets S \setminus IdentifiedCode $
    
    \STATE
    \STATE $MEP \gets EntryPointDetection(S)$
    \STATE \textbf{for each} $a_i \in MEP$ \textbf{do}
        \INDSTATE $C_i \gets RecursiveDisassemble(S, a_i)$
        \INDSTATE $S \gets S \setminus C_i$
    \STATE \textbf{end for}
        
    \STATE
    \STATE \textbf{function} $EntryPointDetection(B):$
        \INDSTATE $ MatchedEntryPoints \ = \left \{ \right \}, Abbreviated\ as\ M$
        \INDSTATE $ JT \gets JumpTableIdentification(B)$
        \INDSTATE $ FU \gets FrameUnwindInformation(B)$
        \INDSTATE $ AF \gets AddressTakenFunctions(B)$
        \INDSTATE $ FE \gets FunctionEntryIdentification(B)$
        \INDSTATE \textbf{for each} $A \in \{JT, FU, AF, FE\}$ \textbf{do}
            \INDSTATE[2] $M \gets M \cup A$
        \INDSTATE \textbf{end for}
        \INDSTATE \textbf{Return} $M$
    \STATE \textbf{end function}

\end{algorithmic} 
\end{algorithm}

\subsection{Algorithm of \MyDisas} \label{sec:DisasAlgo}

Algorithm~\ref{alg:unidirectional} outlines the \MyDisas\ strategy.
As shown in Algorithm~\ref{alg:unidirectional}, the input consists of the stripped binary code $BC$ and the program entry point $EP$.
The algorithm's final output is a superset of embedded data $S$, which includes all embedded data and some code. 
Initially, the entire code section is marked as embedded data superset (Line 1 in Algorithm~\ref{alg:unidirectional}).
Then, we apply the recursive traversal algorithm~\cite{schwarz2002disassembly}, following the control flow from the program entry point to identify the code located on the main paths (Line 2). 
We then exclude the identified code from the superset (Line~3), resulting in a smaller superset.
To further reduce the superset, we conduct multiple additional analyses to uncover missed code entry points, subsequently applying recursive traversal disassembly to these identified entry points (Lines 5-7). During disassembly from each entry point, the identified code is excluded from the superset (Line 9), thereby minimizing the embedded data superset.
Our analyses include examining jump tables, frame unwind information, address-taken functions~\cite{miTrimmer}, and employing function entry identification heuristics~\cite{Egalito} to identify additional code entry points that were not reached by recursive traversal disassembly (Lines~11-21).

\subsection{Performance on Real-World Applications} \label{sec:realworld_apps}

\begin{table} [ht]
	\centering
	\caption{Overhead of \mytool-protected web servers using ApacheBench. The first row presents the request page size,
	         and the data show the additional overhead for each page size.}
	\label{table:WebServers}
	\resizebox{0.47\textwidth}{!}{
		\begin{tabular}{l c c c c c c c}
			\toprule
			Web Server         & 50KB		& 100KB     & 200KB		& 500KB     & 1024KB  & Avg. \\
            \midrule
            Nginx              & 0.11\%		& 0.35\%	& 0.37\%   	& 0.54\%	& 0.38\%  & 0.35\%  \\
			Apache             & 0.44\%		& 0.04\%	& 0.14\%	& 0.33\%	& 0.63\%  & 0.31\% \\
            Lighttpd           & 0.31\%		& 0.43\%	& 0.51\%	& 0.35\%	& 0.34\%  & 0.38\%  \\
			\midrule
			Average			   & 0.29\%     & 0.27\%    & 0.34\%    & 0.41\%    & 0.45\%  & 0.35\%  \\
            \bottomrule
		\end{tabular}	
	    }
\end{table}

\noindent \textbf{Web Servers }
We tested three mainstream web servers: Nginx-1.20.1, Apache-2.4.49, and Lighttpd-1.4.59.
We compile and run the standard and protected web server versions, respectively.
We use ApacheBench~\cite{ApacheBench} to simulate 500 clients to send HTTP requests $100,000$ times asynchronously,
and we record their running time to complete these $100,000$ requests.
To demonstrate the performance impact of {\mytool} when requesting different page sizes,
we tested five different page sizes: 50KB, 100KB, 200KB, 500KB, and 1MB.
As shown in Table~\ref{table:WebServers}, the maximum overhead value is only 0.63\%, and the average overhead is 0.35\%,
and most overhead values are very close to zero.
This indicates that the performance impact of \mytool\ protection on I/O bound web servers is also negligible.

\begin{table} [ht]
	\centering
	\caption{Performance evaluation result for databases.}
	\label{table:Databases}
	\resizebox{0.38\textwidth}{!}{
		\begin{tabular}{l l l c}
			\toprule
			Database        			& Original		& XoM			& Overhead	 \\
            \midrule
            \multirow{2}{*}{MySQL}      & 2132.68/s		& 2119.24/s    	& 0.63\%	 \\
										& 20297.05/s	& 20219.92/s	& 0.38\%	\\
			\midrule
			\multirow{2}{*}{MongoDB}    & 12071.41/s	& 12065.37/s   	& 0.05\%    \\
										& 14745.69/s	& 14735.37/s	& 0.07\%	\\
			\midrule
            \multirow{2}{*}{Redis}      & 143685.12/s   & 143369.01/s   & 0.22\%     \\
										& 139891.09/s	& 139471.42/s	& 0.30\% 	\\
			\midrule
			\multirow{2}{*}{SQLite}     & 650.01/s		& 649.62/s	    & 0.06\%     \\
										& 84901.09/s	& 84807.70/s	& 0.11\% 	\\
			\midrule
			\multirow{2}{*}{Average}    &\multirow{2}{*}{N/A} 	&\multirow{2}{*}{N/A}    & 0.24\%     \\
										& 				& 				& 0.22\% 	\\
            \bottomrule
		\end{tabular}	
	    }
\end{table}

\vspace*{2pt}
\noindent \textbf{Databases }
We tested popular databases: MySQL-8.0.26, MongoDB-4.2.17, Redis-6.2.5, and SQLite-3.36.0.
Unlike ApacheBench, there is no such unified performance benchmark for databases. We have to
run each database with a customized testing suite.
For MySQL, MongoDB and Redis, we use their official benchmarks (sysbench~\cite{sysbench}, mongo-perf~\cite{mongo-perf}, and redis-benchmark~\cite{redis-benchmark})
to evaluate the insertion and selection overhead.
For SQLite, we design a custom benchmark to simulate other three database benchmarks' workloads.
We configure each testing suite as follows:
\begin{enumerate}
	\item MySQL: We configure sysbench with the \emph{complex} workload to perform insertion and selection for $100,000$ rows.
    \item MongoDB: We execute the \emph{simple insert} and \emph{simple query} workload and
				   record how many requests can be completed per second to evaluate the insertion and selection overhead.
	\item Redis: We execute the \emph{SET} and \emph{GET} operation $100,000$ times and record how many requests can be processed per second.
	\item SQLite: We insert $100,000$ rows of random data and select the inserted data by their primary key.
				  We record the execution time and calculate how many requests that SQLite can process per second.
\end{enumerate}

Table~\ref{table:Databases} shows the result of performance evaluation of \mytool\ protection on four widely-used databases.
For each database, the first row shows the insertion performance, and the second row shows the selection performance.
The last column shows the additional overhead incurred by \mytool,
and the last row shows the average overhead for insertion and selection.
As shown in Table~\ref{table:Databases}, the majority of the additional overhead values for databases are almost zero,
and the average overhead is only 0.24\% for insertion and 0.22\% for selection.
These findings prove that the \mytool\ has minimal impact on the runtime overhead of the protected databases.

\subsection{Other Microbenchmark Results} \label{sec:lmbench_unrelated}

Table~\ref{table:MicroProcess}, Table~\ref{table:MicroLocalComm}, and Table~\ref{table:MicroFileVM} show the lmbench results for PXoM-low-correlation kernel operations.

\begin{table} [ht]
	\centering
	\caption{Time for process-related kernel operations in $\mu$s. Smaller is better.}
	\label{table:MicroProcess}
	\resizebox{0.5\textwidth}{!}{
		\begin{tabular}{l | c c c c c c c c}
			\toprule
			\multirow{2}{*}{Kernel}     & Null		& Null		 &\multirow{2}{*}{Stat}   &	Open     & Select   & Signal   & Signal     & sh       \\
			                            & Call      & I/O        &                        & Close    & TCP      & Install  & Handle     & Proc       \\
            \midrule
            Standard                    & 0.09		& 0.14    	 & 0.45                   &	1.23     & 1.64     & 0.14     & 1.02       & 788      \\
			PXoM                        & 0.09	    & 0.14    	 & 0.46                   & 1.21     & 1.62     & 0.14     & 1.03       & 791      \\
			\midrule
            Overhead                    & 0.00\%    & 0.00\%    & 2.22\%                  & 1.65\%   & -1.23\%  & 0.00\%   & 0.98\%     & 0.38\%       \\
            \bottomrule
		\end{tabular}	
	    }
\end{table}

\begin{table} [ht]
	\centering
	\caption{Local communication latencies in $\mu$s. Smaller is better.}
	\label{table:MicroLocalComm}
	\resizebox{0.43\textwidth}{!}{
		\begin{tabular}{c c c c c c}
			\toprule
			\multirow{2}{*}{Kernel}   & Pipe    & AF        & UDP     & TCP     & TCP       \\
			                          &         & UNIX      &         &         & Conn   \\
            \midrule
            Standard                  & 5.25    & 5.22      & 7.71    & 7.80    & 12.08  \\
			PXoM                      & 5.21    & 5.35      & 7.72    & 7.76    & 12.03   \\
			\midrule
            Overhead                  & -0.77\% & 2.49\%    & 0.13\%  & -0.52\% & -0.42\%   \\
            \bottomrule
		\end{tabular}	
	    }
\end{table}

\begin{table} [ht]
	\centering
	\caption{File \& system latencies in $\mu$s. Smaller is better.}

	\label{table:MicroFileVM}
	\resizebox{0.5\textwidth}{!}{
		\begin{tabular}{c c c c c c c}
			\toprule
			\multirow{2}{*}{Kernel}   & \multicolumn{2}{c}{0K File}    & \multicolumn{2}{c}{10K File}    & Mmap      & 100fd      \\ \cmidrule(lr){2-3} \cmidrule(lr){4-5}
			                          & Create   & Delete              & Create     & Delete             & Latency   & Select  \\
            \midrule
            Standard                  & 6.93     & 4.56                & 12.5       & 7.31               & 52.5K     & 0.78  \\
			PXoM                      & 6.98     & 4.53                & 12.4       & 7.39               & 53.1K     & 0.78 \\
			\midrule
            Overhead                  & 0.72\%   & -0.66\%             & -0.81\%    & 1.09\%             & 1.14\%    & 0.00\% \\
            \bottomrule
		\end{tabular}	
	    }
\end{table}

\begin{figure*} [t]
	\centering
	\includegraphics[width=0.95\textwidth]{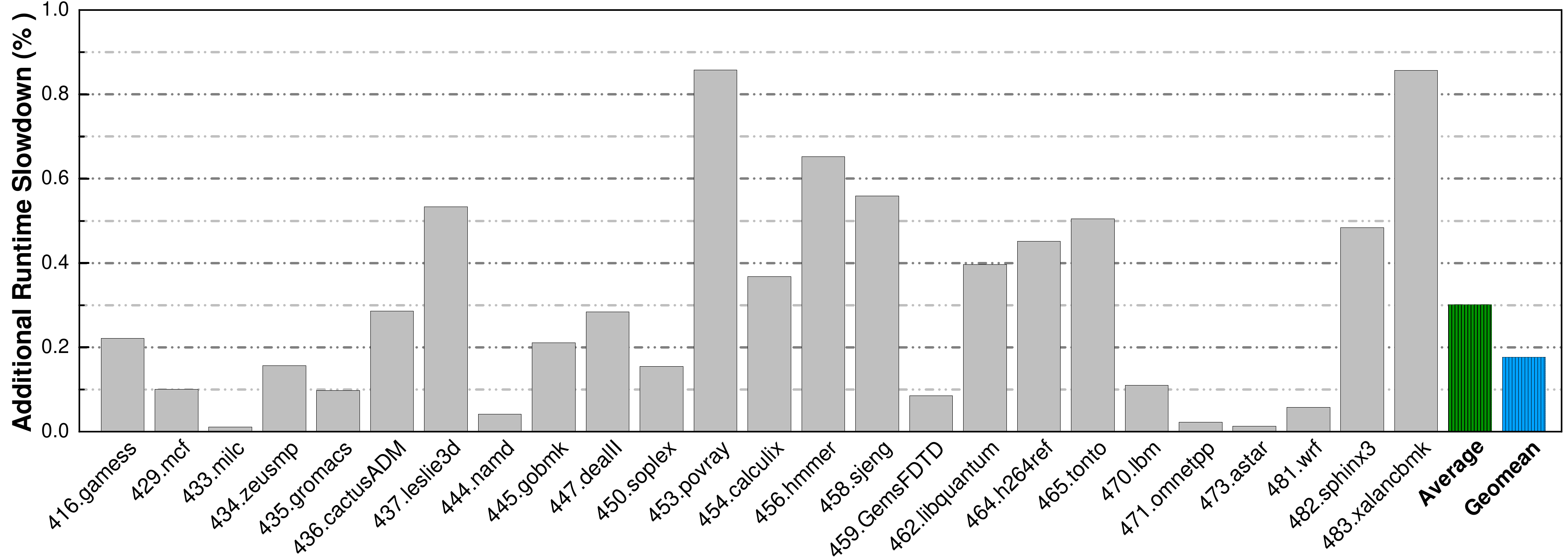}
	\caption{Additional runtime slowdown (\%) of \mytool\ on SPEC CPU 2006 (\emph{ref} workload).}
	\label{fig:SPEC2006}
\end{figure*}

\subsection{Performance Comparison}  \label{sec:perf-compar}
We conducted a separate experiment to compare \mytool\ with other prominent peer tools, including 
XnR~\cite{XnR}, LR$^{2}$~\cite{LR2}, Readactor~\cite{Readactor}, HideM~\cite{HideM}, SECRET~\cite{SECRET}, Heisenbyte~\cite{Heisenbyte}, and Near~\cite{NEAR}.
Like \mytool, they also protect userland programs on x86 platforms.
However, we encountered a major challenge in the current inability to reproduce or replicate previous XoM results, which is an issue that unfortunately plagues the security field.
This problem is particularly evident in the area of XoM, as none of the XoM papers listed in Table~\ref{table:XOMComparison} or kernel-level XoM papers~\cite{KHide,KRX} have released their tools publicly.
In light of this, we evaluated the additional overhead caused by \mytool\ on SPEC 2006, and compared the results with reported performance data in their papers.

Figure~\ref{fig:SPEC2006} shows the additional overhead of \mytool's protection for SPEC CPU 2006.
Similar to SPEC CPU 2017's performance data, \texttt{453.povray} and  \texttt{483.xalancbmk} exhibit relatively high overhead.
Overall, \mytool\ incurs smaller runtime overhead on SPEC CPU 2006 than SPEC CPU 2017; 
the average and geometric mean overhead of SPEC CPU 2006 benchmarks are 0.30\% and 0.18\%, respectively
We attribute this difference to the larger and more complex workloads executed by SPEC CPU 2017.
SPEC CPU 2006 is no longer sufficient to meet the load demands of modern CPUs.

\begin{figure} [t]
  \centering
  \includegraphics[width=0.43\textwidth]{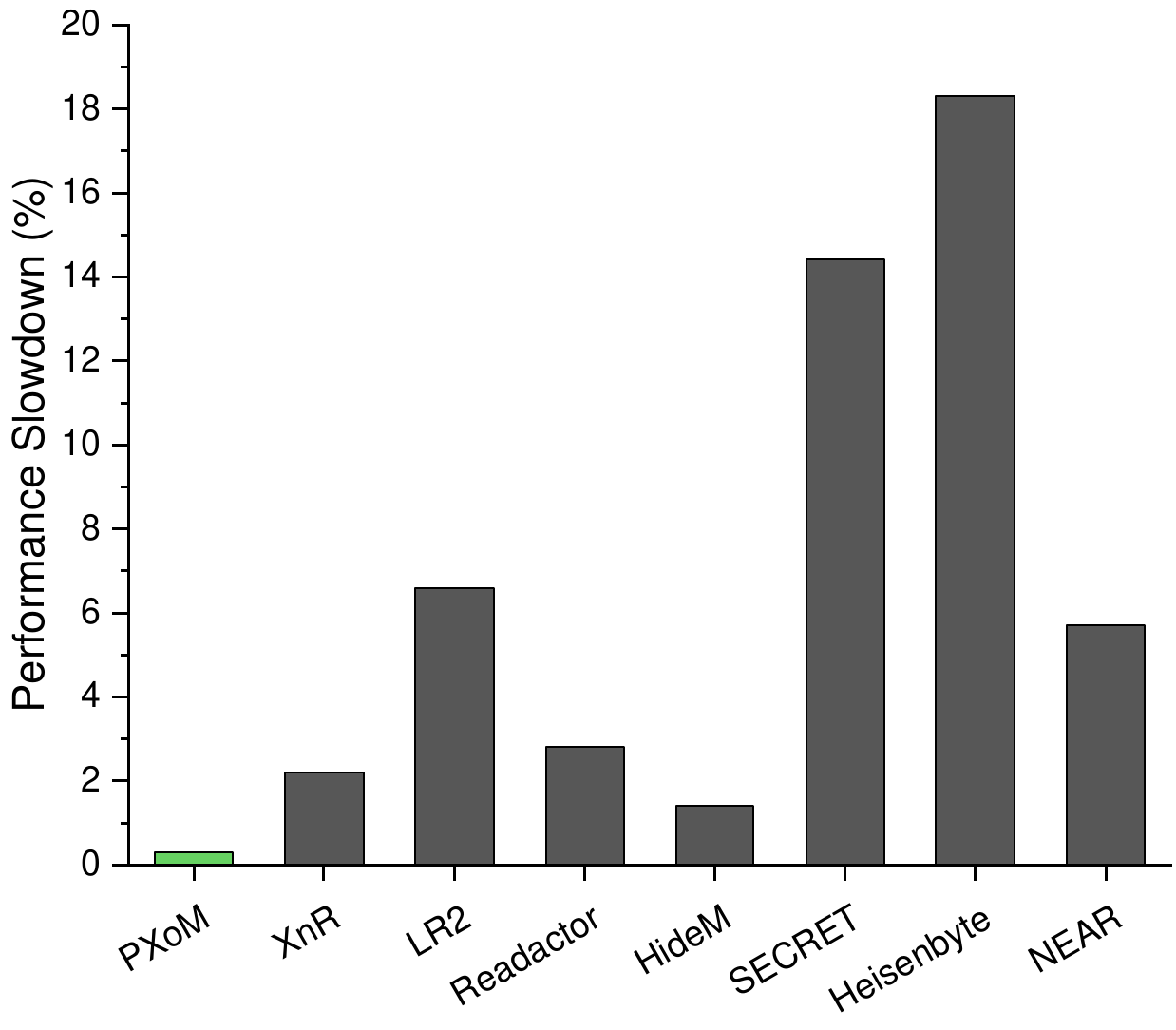}
  \caption{Performance comparison with other XoM implementations. Although some papers such as Readactor~\cite{Readactor} have other security modules (e.g., fine-grained code randomization), here we only present XoM's performance data. }
  \label{fig:SPEC_Compare}
\end{figure}

Figure~\ref{fig:SPEC_Compare} shows the performance comparison result.
Heisenbyte has a significantly higher overhead (18.3\%) compared to other approaches,
which we attribute to its intricate legitimate data-in-code read process.
NEAR reduces the overhead to 5.7\% by simplifying this process.
Readactor achieved XoM through  hardware virtualization, which incurs a 5.8\% performance penalty.
LR$^{2}$ causes a medium overhead (6.6\%) due to its software-fault isolation.
XnR and HideM reveal 2.2\% and 1.4\% extra slowdown, respectively.
In contrast, \mytool\ exhibits the lowest overhead (0.30\%) among all XoM prototypes in Figure~\ref{fig:SPEC_Compare}.

\subsection{Case Studies for Security Evaluation} \label{sec:casestudy}

\noindent \textbf{Exploit Memory Disclosure Vulnerability }
We use the CVE-2013-2028~\cite{CVE-2013-2028}, a Nginx arbitrary memory disclosure vulnerability, to showcase \mytool's effectiveness.
This vulnerability is a potent stack overflow that enables an attacker to carry out arbitrary memory reads.
We apply \mytool\ to a vulnerable binary version of Nginx and run it as a web server.
After that, we modify the JIT-ROP attack framework, jitrop-native~\cite{jirop-native}, to specifically adapt it to the CVE-2013-2028, and use it to trigger this vulnerability and dynamically search for gadgets.
The attack is detected and prevented when the exploit tries to reveal the first code byte,
indicating that \mytool\ is capable of safeguarding Nginx from memory disclosure.

\vspace*{2pt}
\noindent \textbf{Construct WRPKRU Gadgets}
The value of PKRU can be changed using the \emph{WRPKRU} instruction at the user level.
Intuitively, if this instruction is located by attackers in executable data blocks, they can use it to regain read permission for the code pages.
In order to successfully change the permission of a page group using the WRPKRU instruction, four operations must be performed:
1) storing the permission value to EAX; 2) writing zero to ECX; 3) writing zero to EDX; and 4) executing WRPKRU.
Please note that when executing the WRPKRU instruction, EAX stores the permission value for all page groups that will later be written into PKRU,
and the values of ECX and EDX must be zero to avoid a general-protection exception (\#GP).
Completing these operations may require more than four gadgets.
For instance, in OpenSSL, only \texttt{XCHG} instructions can change \texttt{EAX}'s value,  such as \texttt{XCHG EDI, EAX}.
Thus, an additional gadget is necessary to write the permission value to \texttt{EDI}, which can then be swapped with EAX using \texttt{XCHG}.
To find gadgets capable of completing these operations, we conduct a gadget search on all binaries' executable data in the Pang et al.'s data set~\cite{GroundTruth}.
The dataset revealed that only 23 binaries' executable data contain gadgets that can complete one or two operations, but no binary gadgets that can complete all four operations.
Interestingly, even treating each byte as an opcode, we only found 26 WRPKRU instructions in the $\sim20GB$ dataset, and none of them was in the executable data areas.
This rarity of the byte sequence of WRPKRU (\texttt{0F 01 EF}) in the compiled binary could explain the difficulty of finding gadgets capable of performing all four operations.

\vspace*{2pt}
\noindent \textbf{Attacker-Controllable Syscalls }
Despite the fact that attackers may attempt to leverage system calls (e.g., \emph{mprotect} and \emph{execve}) to conduct their second-stage attack, 
thereby circumventing the necessity for Turing-complete gadgets and minimizing gadget requirements, 
they still necessitate multiple gadgets to manipulate the parameters of these system calls.
Johannesmeyer et al.~\cite{usenix24dataonly} enumerated twelve system calls that could potentially be exploited to implement such attacks. 
We conducted an additional experiment to search for gadgets capable of manipulating the parameters of these twelve system calls 
within the embedded data regions of the dataset presented in Table~\ref{table:DisassemblyResult}, and no such gadgets were discovered. 

\begin{figure} %[t]
  \centering
  \includegraphics[width=0.43\textwidth]{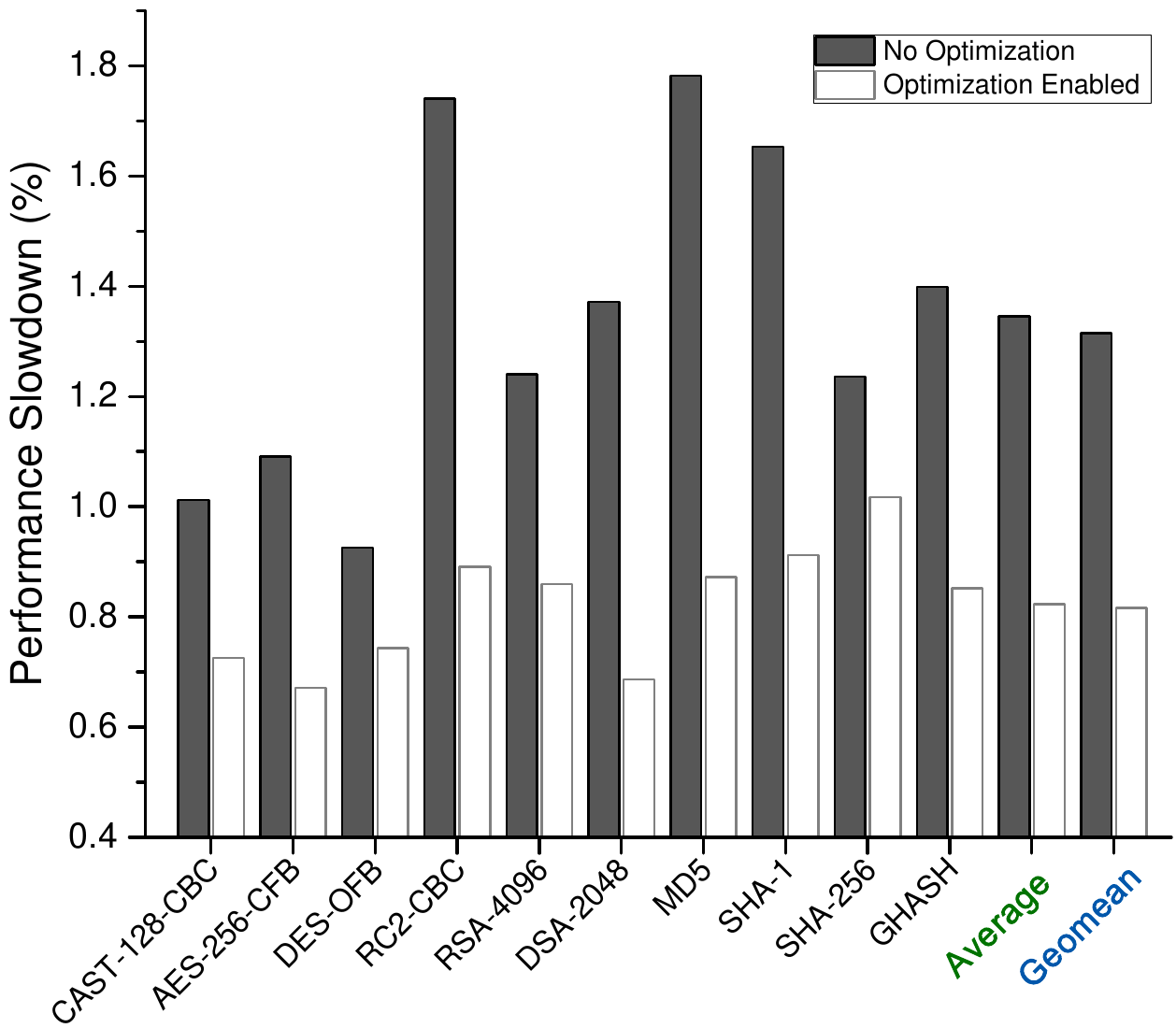}
  \caption{Optimization effect measurement on OpenSSL.}
  \label{fig:openssl}
\end{figure}

\subsection{High-Frequency Data Read Optimization} \label{sec:optimization} %High-Frequency Legitimate Embedded Data Reads Optimization

In the scrutiny of a read request, \mytool\ faces potential performance bottlenecks when iteratively navigating an extensive list of embedded data, 
a concern exacerbated in programs featuring a large number of code-data interleaving cases. 
Our empirical observations of real-world programs reveal that, in scenarios where data-in-code reads occur frequently, the read targets tend to cluster around a confined subset of embedded data.
For example, during the execution of AES-256-CBC encryption in OpenSSL, the program exclusively accesses a mere $15$ out of $8142$ embedded data blocks.
This motivates us to implement a cache-like optimization policy that accelerates the legitimacy determination of frequently-read embedded data.
We segregate the high-frequency read embedded data into a separate \emph{optimization list}, while preserving other embedded data in a \emph{regular list}.
We ensure that the optimization list remains concise, prioritizing its iteration when validating the legitimacy of data-in-code reads.
Our strategy for creating the optimization list encompasses both static and dynamic policies. Under the static policy, we consider the frequency of references to embedded data, 
relegating embedded data with over 10 references to the optimization list. Meanwhile, the dynamic policy involves real-time monitoring within the exception handler, recording the frequency of reads for embedded data. 
Data surpassing 100 reads dynamically qualifies for inclusion in the optimization list. 
We have empirically determined the threshold of 10 references and 100 reads to achieve the optimal performance.
\mytool\ turns on this optimization policy by default.

From Figure~\ref{fig:ReadLatency}, it is evident that as the size of the embedded data list expands, the time required for data-in-code reads also increases.
Our analysis indicates that programs that engage in frequent data-in-code reads usually target a small subset of embedded data blocks. 
To alleviate this overhead, our cache-like optimization policy comes into play, effectively reducing the size of the embedded data list when programs perform frequent data-in-code reads.
We use OpenSSL to evaluate the effectiveness of the optimization policy in reducing overhead for high-frequency data-in-code reads.
Without our optimization policy, each read-legality check will take extra time to traverse the long embedded data list.
With our optimization policy, the performance can be significantly improved.
The comparative experiment results on OpenSSL, with and without our optimization policy, are shown in Figure~\ref{fig:openssl}.
The average overhead, without our optimization policy, was 1.35\%, while the geometric mean was 1.31\%. In contrast, with our optimization enabled,
the average overhead was reduced to 0.82\%, and the geometric mean was lowered to 0.81\%.
We also conducted an optimization policy evaluation on BoringSSL and FFmpeg, and the conclusions were consistent with those from OpenSSL.

\subsection{Case Studies of Embedded Data in COTS Binaries} \label{sec:COTS_CaseStudy}

Through empirical study, we categorized the embedded data into the following four types:
\begin{enumerate}
    \item Embedded Constants: Independent constants dispersed throughout the program, each referenced separately. These constants can include integers, floating-point numbers, or other data structures.
    \item Embedded Arrays: Groups of constants organized into arrays, where each element is accessed using an ``\texttt{array pointer + index}.''
    \item Embedded Strings: Strings embedded directly within the code section.
    \item Jump Tables: Tables that store target addresses for switch-case structures.
\end{enumerate}

\subsubsection{Embedded Constants}

Using Skype version 8.129.0.202 as an example, Figure~\ref{disas:EmbeddedNumbers} illustrates the embedded constants within the binary file \textit{skypeforlinux}. In lines 1 and 3, two 128-bit integers are embedded, with the \texttt{paddd} instruction being used in lines 10 and 12 to add them to the value in the \texttt{xmm0} register. Similarly, in lines 5 and 7, two 256-bit integers are embedded, which are then used in lines 14 and 17.

\begin{figure} [ht]
\begin{lstlisting}[style=NewASMStyle]
.text:1A913A0 xmmword_1A913A0 xmmword 
              3000000020000000100000000h
.text:1A913B0 xmmword_1A913B0 xmmword 
              4000000040000000400000004h
.text:1A913C0 ymmword_1A913C0 ymmword 
              0000000002000000040...(256 Bits)
.text:1A913E0 ymmword_1A913E0 ymmword
              0800000008000000080...(256 Bits)
...
.text:1A91C3F paddd xmm0, cs:xmmword_1A913A0
...
.text:1A91CDB paddd xmm0, cs:xmmword_1A913B0
...
.text:1A92631 vpaddd ymm4, ymm4,
                     cs:ymmword_1A913C0
...
.text:1A926AE vpaddd ymm4, ymm4, 
                     cs:ymmword_1A913E0
...
\end{lstlisting}
\caption{Embedded constants in the \textit{skypeforlinux} of Skype.}
\label{disas:EmbeddedNumbers}
\end{figure}

\subsubsection{Embedded Arrays}

Unlike embedded constants, where each constant has its own reference, embedded arrays group constants together, with each element accessed via an array pointer. For example, in the main binary \textit{resolve} of DaVinci Resolve (version 19.0.1), there is an embedded array, as shown in Figure~\ref{disas:EmbeddedArrays}. In line 9, an array of 4,160 bytes is embedded, and in line 1, its reference is loaded into the \texttt{rbp} register. Lines 3, 5, and 6 show how values from the array are accessed using the array pointer stored in the \texttt{rbp} register, with the \texttt{rsi} register serving as the index. The retrieved values are then loaded into the \texttt{r8} and \texttt{r9} registers.

\begin{figure} [ht]
\begin{lstlisting}[style=NewASMStyle]
.text:A5DEFF2 lea rbp, qword_A5DF8C0;Array Ref
...           
.text:A5DF0A0 mov r8, [rbp+rsi*8+1000h]
...           ; Get values from an array
.text:A5DF0C6 xor r8, [rbp+rsi*8+0]
.text:A5DF0CB mov r9, [rbp+rdi*8+7]
...
qword_A5DF8C0:   ; An array of 4160 bytes
.text:A5DF8C0 dq 2 dup(0D83078C018601818h),
                 2 dup(2646AF05238C2323h)
.text:A5DF8E0 dq 2 dup(0B891F97EC63FC6C6h), 
                 2 dup(0FBCD6F13E887E8E8h)
.text:A5DF900 dq 2 dup(0CB13A14C87268787h), 
                 2 dup(116D62A9B8DAB8B8h)
...              
\end{lstlisting}
\caption{An embedded array in the \textit{resolve} of DaVinci Resolve.}
\label{disas:EmbeddedArrays}
\end{figure}

\subsubsection{Embedded Strings}

Embedded strings are a specific type of embedded constant, characterized by their variable size and termination with a 0x00 byte. Figure~\ref{disas:EmbeddedStr} provides an example of embedded strings found in the \textit{teamviewerd} daemon of TeamViewer (version 15.58.4). In lines 6, 7, and 8, three strings are embedded, with a reference to the string embedded in line 6 made in line 1.

\begin{figure} [ht]
\begin{lstlisting}[style=NewASMStyle]
.text:A748F4  lea rax, aRc48xInt ;String Ref
.text:A748FB  mov edx, dword cs:qword_16781C0
.text:A74901  bt  edx, 14h
.text:A74905  jb  short loc_1F3853
...
.text:A74940 aRc48xInt  db 'rc4(8x,int)',0 
.text:A7494C aRc48xChar db 'rc4(8x,char)',0
.text:A74959 aRc416xInt db 'rc4(16x,int)',0
...
\end{lstlisting}
\caption{Embedded strings in the \textit{teamviewerd} daemon of TeamViewer.}
\label{disas:EmbeddedStr}
\end{figure}

\begin{figure} [ht]
\begin{lstlisting}[style=NewASMStyle]
.text:7D642  lea  rsi, jpt_7D666
.text:7D649  mov  r10, rdi
.text:7D64C  neg  r10
.text:7D64F  add  r10, 40h
.text:7D653  and  r10, 3Fh  ;Switch 64 cases
...
.text:7D666  jmp  rsi       ;Switch Jump
...
jpt_7D666:
.text:7D8C0  dq 7D8C0h - offset loc_7D726,
                7D8C0h - offset loc_7D723
.text:7D8D0  dq 7D8C0h - offset loc_7D72B,
                7D8C0h - offset loc_7D734
.text:7D8E0  dq 7D8C0h - offset loc_7D741,
                7D8C0h - offset loc_7D749
.text:7D8F0  dq 7D8C0h - offset loc_7D749,
                7D8C0h - offset loc_7D749
...
\end{lstlisting}
\caption{A jump table in the \textit{dpcpp} of Intel DPC++.}
\label{disas:JmpTable}
\end{figure}

\subsubsection{Jump Tables}

A jump table is a specialized type of embedded array and one of the most common data structures embedded in code. It stores the target addresses for switch-case structures. In Position Independent Code (PIC), the jump table holds the offset between the target code and the jump instruction, while in non-PIC code, it stores the absolute address of the target code.
Although compilers like GCC and LLVM typically place jump tables in the data segment, some compilers, such as Intel's C++ compiler, prefer to embed jump tables closer to the code that uses them. This approach reduces the likelihood of cache misses, thereby improving program performance. In contrast, placing the jump table in a distant data segment can increase the frequency of cache misses.

Figure~\ref{disas:JmpTable} provides an example of a jump table in the \textit{dpcpp} binary of Intel DPC++ (version 2.1.79). At line 10, a switch structure with 64 cases is defined. Since this binary is a position-independent executable (PIE), the jump table starting at line 10 stores offsets of the target addresses relative to line 7. In line 1, the address of the jump table is loaded into the \texttt{rsi} register, and after performing a series of calculations based on the index value, the target address is determined. Finally, in line 7, the program jumps to the calculated target address.

%% file: A2_artifact_appendix.tex
% Artifact Appendix template for the NDSS Artifact Evaluation
% version 1.0 (20230620)

% remove the following block when merging the appendix with the camera-ready full paper
%%%

\renewcommand{\appendixname}{Artifact Appendix}
\appendix

\subsection{Description \& Requirements}

In our paper, we present PXoM, a hardware-assisted approach to retrofitting XoM (Execute-only Memory) for stripped binaries, without the need for relocating embedded data.
PXoM is a comprehensive system that includes a full-stack toolchain, from user-level applications to a custom kernel, designed to provide XoM protection for programs while ensuring compatibility with legitimate embedded data reads within code sections, all without the need for relocating embedded data.

In this artifact, we provide the following:
\begin{enumerate}
    \item Virtual Machine (PXoM\_Artifact.ova): An out-of-the-box (OOB) virtual machine with a customized system kernel and user-space toolchain pre-deployed for easy access to PXoM. This VM offers a convenient way to quickly start testing PXoM on various programs and reproducing the evaluations presented in this paper.
    \item Source Code (PXoM\_Artifact-0.1.tar.gz): The source code for the PXoM kernel, user-space toolchain, and all the experiments described in this artifact.
    \item Documentation (PXoM\_Artifact.pdf): Detailed instructions on how to use the PXoM virtual machine and the workflow for conducting the experiments.
\end{enumerate}

In this artifact appendix, we will outline the hardware and software requirements for PXoM, the steps to install the virtual machine, the major claims from our paper, and the experimental workflows.

\noindent \textbf{How to access }

PXoM Virtual Machine: \href{https://zenodo.org/records/13892220}{10.5281/zenodo.13892220}

Source Code: \href{https://zenodo.org/records/14251050}{10.5281/zenodo.14251050}

PXoM Artifact Documentation: \href{https://zenodo.org/records/14251155}{10.5281/zenodo.14251155}

\vspace*{2pt}
\noindent \textbf{Hardware dependencies}

The only required hardware feature is MPK, which is supported by the following CPUs:
\begin{enumerate}
    \item Intel{\textregistered} Desktop CPUs, Comet Lake (10th Gen Core{\texttrademark{}}) and later;
    \item Intel{\textregistered} Server CPUs, Xeon{\textregistered} Skylake and later;
    \item AMD Desktop CPUs, Ryzen{\texttrademark{}} 5000 and later;
    \item AMD Server CPUs, EPYC{\texttrademark{}} Milan (7003 Series) and later.
\end{enumerate}

\vspace*{2pt}
\noindent \textbf{Software dependencies} 

There are two options for deploying the PXoM VM:

\begin{enumerate}
    \item On a host running Ubuntu 22.04 or later, you can import the PXoM virtual machine using VMWare Workstation Pro 17.6.0 or higher. Please note that MPK is not supported on Windows hosts, so VMWare Workstation Pro must be installed on a Linux-based host. 
    \item Directly import the PXoM virtual machine on a machine running ESXi 8.0 Update 3 or later.
\end{enumerate}

\noindent \textbf{Benchmarks} 

Most of the benchmarks are included with the PXoM VM image. However, we have excluded Pang et al.'s dataset from the image due to its large size ($\sim$56GB after decompression), which would make the image excessively bloated. You can obtain Pang et al.'s dataset from their \href{https://github.com/junxzm1990/x86-sok}{Github repository}.
Please refer to the instructions in the artifact documentation (PXoM\_Artifact.pdf) before obtaining the dataset.

\subsection{Artifact Installation \& Configuration}

The only installation step is to import the PXoM virtual machine (PXoM\_Artifact.ova) into VMWare Workstation Pro or VMWare ESXi. All the experiments from our paper can be conducted within this virtual machine.

For instructions on how to import the OVA file into VMWare Workstation Pro and VMWare ESXi, please refer to the \href{https://docs.vmware.com/en/VMware-Workstation-Pro/17/com.vmware.ws.using.doc/GUID-DDCBE9C0-0EC9-4D09-8042-18436DA62F7A.html}{VMware Workstation documentation} and \href{https://docs.vmware.com/en/VMware-vSphere/7.0/com.vmware.vsphere.hostclient.doc/GUID-8ABDB2E1-DDBF-40E3-8ED6-DC857783E3E3.html}{VMware ESXi documentation}.
After importing the PXoM virtual machine, you can adjust its memory size and the number of CPUs. We recommend allocating more than 32GB of memory and assigning more than 10 cores.

\noindent \textbf{Optional:} To compile the PXoM kernel on a bare-metal machine, please follow the same process as you would for the standard Linux kernel. For example, begin with the instructions starting at Step 3 in this guide: \url{https://phoenixnap.com/kb/build-linux-kernel}.

\subsection{Major Claims}

Our paper makes two major claims:
\begin{itemize}
    \item (C1): \textsc{PXoM} can protect stripped binaries from JIT-ROP attacks while allowing legitimate embedded data reads, without requiring relocating embedded data. This is demonstrated through experiments (E1) and (E2).
    \item (C2): \textsc{PXoM} introduces negligible performance overhead. This is validated by experiments conducted on lmbench (E3), SPEC CPU 2017 (E4), Webservers (E5), and Databases (E6).
\end{itemize}

\subsection{Evaluation}

The experiments are divided into six parts:
(E1): JIT-ROP Defense Demonstration; (E2): Disassembly Result Evaluation; (E3): lmbench; (E4): SPEC CPU 2017; (E5): Web Servers; and (E6): Databases.
E1 and E2 support C1, while E3$\sim$E6 support C2. We provide instructions to reproduce the experiments described in our paper; however, we do not claim the ``reproduced'' badge for two reasons:

\begin{enumerate}
    \item The dataset for E2 is too large, requiring 5 to 6 days to fully evaluate the entire dataset.
    \item The virtual machine provides the most reliable environment to ensure PXoM functions correctly by masking hardware differences, which ensures proper operation across various devices. However, virtualization may lead to inaccurate performance evaluation results. Moreover, Intel's hybrid architecture of E-Cores and P-Cores can further amplify experimental inaccuracies.
\end{enumerate}

The \texttt{$\sim$/PXoM\_Artifact/experiments} folder within the virtual machine contains all the experiments. Please conduct each experiment in its corresponding folder.

\subsubsection{\textbf{Experiment (E1)}}
[JIT-ROP Defense Demonstration] [20 human-minutes + 5 compute-minutes]: Demonstrating how PXoM protects programs from JIT-ROP attacks while allowing legitimate embedded data reads.

[\textit{Workflow}] 

1. Run the vulnerable demo server:
\vspace*{-2pt}
\begin{lstlisting}[style=BashInputStyle]
    >  sudo  ./cipher_helper
\end{lstlisting}
\vspace*{-4pt}

2. Exploit the vulnerable server:
\vspace*{-2pt}
\begin{lstlisting}[style=BashInputStyle]
    >  python  exploit.py
\end{lstlisting}
\vspace*{-4pt}

3. Protect the vulnerable server using the PXoM toolchain:
\vspace*{-2pt}
\begin{lstlisting}[style=BashInputStyle]
    >  pxom  protect  -i  cipher_server  -o  cipher_server.xom
\end{lstlisting}
\vspace*{-4pt}

4. Run the protected server and attempt the exploit again:
\vspace*{-2pt}
\begin{lstlisting}[style=BashInputStyle]
    >  sudo  ./cipher_helper.xom
    >  python  exploit.py
\end{lstlisting}
\vspace*{-4pt}

5. The exploitation will fail, and you can check the kernel log for details:
\vspace*{-2pt}
\begin{lstlisting}[style=BashInputStyle]
    >  sudo  dmesg
\end{lstlisting}
\vspace*{-4pt}

\subsubsection{\textbf{Experiment (E2)}}
[Disassembly Result Evaluation] [15 human-minutes + 10 compute-minutes (5$\sim$6 days for the entire dataset)]: Evaluating the effectiveness of our disassembly strategy.

[\textit{Workflow}]

1. Protect the program and print the embedded data list:
\vspace*{-2pt}
\begin{lstlisting}[style=BashInputStyle]
    >  pxom  protect  -i  openssl_O3  -o  openssl_O3.xom
    >  pxom  print  -i  openssl_O3.xom  >  xom_edata
\end{lstlisting}
\vspace*{-4pt}

2. Extract the ground truth using Pang et al.'s toolchain:
\vspace*{-2pt}
\begin{lstlisting}[style=BashInputStyle]
    >  objcopy  --dump-section  .rand=tmp_gt.gz  openssl_O3
&&  gzip  -d  tmp_gt.gz
    >  python  extract_gt/extract_edata.py  -b  openssl_O3  -m
tmp_gt  -o  tmp_pb  -L  gt_edata
\end{lstlisting}
\vspace*{-4pt}

3. Compare the disassembly results with the ground truth:
\begin{lstlisting}[style=BashInputStyle]
    >  python  compare_results.py  openssl_O3  xom_edata
gt_edata
\end{lstlisting}
\vspace*{-4pt}

[\textit{Results}] Code Coverage and Overall Coverage for binaries.

\subsubsection{\textbf{Experiment (E3)}}
[lmbench] [10 human-minutes + 20 compute-minutes]: Evaluating the performance overhead of PXoM's kernel modifications.

[\textit{Workflow}]

1. Run lmbench:
\begin{lstlisting}[style=BashInputStyle]
    >  make results
\end{lstlisting}
\vspace*{-4pt}

2. Compare the results with the baseline results.

[\textit{Results}] Performance overhead of kernel modification.

\subsubsection{\textbf{Experiment (E4)}}
[SPEC CPU 2017] [20 human-minutes + 6 compute-hours]: Evaluating the performance overhead of PXoM's protection on compute-intensive programs.

[\textit{Workflow}]

1. Run the standard SPEC CPU 2017:
\vspace*{-2pt}
\begin{lstlisting}[style=BashInputStyle]
    >  ./run_standard.sh
\end{lstlisting}
\vspace*{-4pt}

2. Run the PXoM-protected version of SPEC CPU 2017:
\vspace*{-2pt}
\begin{lstlisting}[style=BashInputStyle]
    >  ./run_pxom.sh
\end{lstlisting}
\vspace*{-4pt}

3. Compare the results:
\vspace*{-2pt}
\begin{lstlisting}[style=BashInputStyle]
    >  python  compare.py
\end{lstlisting}
\vspace*{-4pt}

[\textit{Results}] Performance overhead of PXoM on protecting SPEC CPU 2017

\subsubsection{\textbf{Experiment (E5)}}
[Web Servers] [15 human-minutes + 10 compute-minutes]: Evaluating the performance overhead of PXoM on protecting web servers.

[\textit{Workflow}]

For each web server, the workflow is the same:

1. Start the standard version of the web server:
\vspace*{-2pt}
\begin{lstlisting}[style=BashInputStyle]
    >  sudo  ./start_standard
\end{lstlisting}
\vspace*{-4pt}

2. Obtain the baseline runtime:
\vspace*{-2pt}
\begin{lstlisting}[style=BashInputStyle]
    >  python  run_standard.py
\end{lstlisting}
\vspace*{-4pt}

3. Stop the standard version of the web server:
\vspace*{-2pt}
\begin{lstlisting}[style=BashInputStyle]
    >  sudo  ./stop_standard
\end{lstlisting}
\vspace*{-4pt}

4. Start the PXoM-protected web server:
\vspace*{-2pt}
\begin{lstlisting}[style=BashInputStyle]
    >  sudo  ./start_pxom
\end{lstlisting}
\vspace*{-4pt}

5. Run the PXoM-protected tests:
\vspace*{-2pt}
\begin{lstlisting}[style=BashInputStyle]
    >  python  run_pxom.py
\end{lstlisting}
\vspace*{-4pt}

6. Stop the PXoM-protected web server:
\vspace*{-2pt}
\begin{lstlisting}[style=BashInputStyle]
    >  sudo  ./stop_pxom
\end{lstlisting}
\vspace*{-4pt}

7. Compare the results:
\vspace*{-2pt}
\begin{lstlisting}[style=BashInputStyle]
    >  python  compare_results.py
\end{lstlisting}
\vspace*{-4pt}

[\textit{Results}] Performance overhead of PXoM on protecting each web server.

\subsubsection{\textbf{Experiment (E6)}}
[Databases] [20 human-minutes + 60 compute-minutes]: Evaluating the performance overhead of PXoM on protecting databases.

[\textit{Workflow}]

For MySQL, MongoDB, and Redis:

1. Start the standard version of the database:
\vspace*{-2pt}
\begin{lstlisting}[style=BashInputStyle]
    >  ./start_standard
\end{lstlisting}
\vspace*{-4pt}

2. Obtain the baseline performance data:
\vspace*{-2pt}
\begin{lstlisting}[style=BashInputStyle]
    >  ./run_standard
\end{lstlisting}
\vspace*{-4pt}

3. Start the PXoM-protected database:
\vspace*{-2pt}
\begin{lstlisting}[style=BashInputStyle]
    >  ./start_pxom
\end{lstlisting}
\vspace*{-4pt}

4. Run the PXoM-protected tests:
\vspace*{-2pt}
\begin{lstlisting}[style=BashInputStyle]
    >  ./run_pxom
\end{lstlisting}
\vspace*{-4pt}

5. Compare the results:
\vspace*{-2pt}
\begin{lstlisting}[style=BashInputStyle]
    >  python  compare_results.py
\end{lstlisting}
\vspace*{-4pt}

For SQLite:

1. Obtain the baseline performance data:
\vspace*{-2pt}
\begin{lstlisting}[style=BashInputStyle]
    >  ./run_standard
\end{lstlisting}
\vspace*{-4pt}

2. Run the PXoM-protected tests:
\vspace*{-2pt}
\begin{lstlisting}[style=BashInputStyle]
    >  ./run_pxom
\end{lstlisting}
\vspace*{-4pt}

3. Compare the results:
\vspace*{-2pt}
\begin{lstlisting}[style=BashInputStyle]
    >  python  compare_results.py
\end{lstlisting}
\vspace*{-4pt}

[\textit{Results}] Performance overhead of PXoM on protecting each database.

%% file: PXoM.bbl
\begin{thebibliography}{10}

\bibitem{Szekeres13}
L\'{a}szl\'{o} Szekeres, Mathias Payer, Tao Wei, and Dawn Song.
\newblock {SoK: Eternal War in Memory}.
\newblock In {\em Proceedings of the 34th IEEE Symposium on Security and Privacy (S\&P'13)}, 2013.

\bibitem{Kuznetzov14}
Volodymyr Kuznetzov, L{\'a}szl{\'o} Szekeres, Mathias Payer, George Candea, R~Sekar, and Dawn Song.
\newblock {Code-Pointer Integrity}.
\newblock In {\em Proceedings of the 11th USENIX Symposium on Operating Systems Design and Implementation (OSDI'14)}, 2014.

\bibitem{Hurdle20}
Christian DeLozier, Kavya Lakshminarayanan, Gilles Pokam, and Joseph Devietti.
\newblock {Hurdle: Securing Jump Instructions Against Code Reuse Attacks}.
\newblock In {\em Proceedings of the 25th International Conference on Architectural Support for Programming Languages and Operating Systems (ASPLOS'20)}, 2020.

\bibitem{PIBE21}
Victor Duta, Cristiano Giuffrida, Herbert Bos, and Erik van~der Kouwe.
\newblock {PIBE: Practical Kernel Control-Flow Hardening with Profile-Guided Indirect Branch Elimination}.
\newblock In {\em Proceedings of the 26th ACM International Conference on Architectural Support for Programming Languages and Operating Systems (ASPLOS'21)}, 2021.

\bibitem{ViK22}
Haehyun Cho, Jinbum Park, Adam Oest, Tiffany Bao, Ruoyu Wang, Yan Shoshitaishvili, Adam Doup\'{e}, and Gail-Joon Ahn.
\newblock {ViK: Practical Mitigation of Temporal Memory Safety Violations through Object ID Inspection}.
\newblock In {\em Proceedings of the 27th ACM International Conference on Architectural Support for Programming Languages and Operating Systems (ASPLOS'22)}, 2022.

\bibitem{miTrimmer}
Haotian Zhang, Mengfei Ren, Yu~Lei, and Jiang Ming.
\newblock {One Size Does Not Fit All: Security Hardening of MIPS Embedded Systems via Static Binary Debloating for Shared Libraries}.
\newblock In {\em Proceedings of the 27th International Conference on Architectural Support for Programming Languages and Operating Systems (ASPLOS'22)}, 2022.

\bibitem{ROP}
Hovav Shacham.
\newblock {The Geometry of Innocent Flesh on the Bone: Return-into-libc without Function Calls (on the x86)}.
\newblock In {\em Proceedings of the 14th ACM Conference on Computer and Communications Security (CCS'07)}, 2007.

\bibitem{Oxymoron}
Michael Backes and Stefan N{\"u}rnberger.
\newblock {Oxymoron: Making Fine-Grained Memory Randomization Practical by Allowing Code Sharing}.
\newblock In {\em Proceedings of the 23rd USENIX Conference on Security Symposium (USENIX Security'14)}, 2014.

\bibitem{Isomeron}
Lucas Davi, Christopher Liebchen, Ahmad-Reza Sadeghi, Kevin~Z Snow, and Fabian Monrose.
\newblock {Isomeron: Code Randomization Resilient to (Just-In-Time) Return-Oriented Programming}.
\newblock In {\em Proceedings of the 22nd Annual Network and Distributed System Security Symposium (NDSS'15)}, 2015.

\bibitem{ASLR}
PaX Team.
\newblock {Address Space Layout Randomization (ASLR)}.
\newblock \url{https://pax.grsecurity.net/docs/aslr.txt}, 2003.

\bibitem{AddressObfuscation}
Sandeep Bhatkar, Daniel~C DuVarney, and Ron Sekar.
\newblock {Address Obfuscation: An Efficient Approach to Combat a Broad Range of Memory Error Exploits}.
\newblock In {\em Proceedings of the 12th Conference on USENIX Security Symposium (USENIX Security'03)}, 2003.

\bibitem{bhatkar2005efficient}
Sandeep Bhatkar, Daniel~C DuVarney, and R~Sekar.
\newblock {Efficient Techniques for Comprehensive Protection from Memory Error Exploits}.
\newblock In {\em Proceedings of the 14th Conference on USENIX Security Symposium (USENIX Security'05)}, 2005.

\bibitem{giuffrida2012enhanced}
Cristiano Giuffrida, Anton Kuijsten, and Andrew~S Tanenbaum.
\newblock {Enhanced Operating System Security Through Efficient and Fine-grained Address Space Randomization}.
\newblock In {\em Proceedings of the 21st Conference on USENIX Security Symposium (USENIX Security'12)}, 2012.

\bibitem{hiser2012ilr}
Jason Hiser, Anh Nguyen-Tuong, Michele Co, Matthew Hall, and Jack~W Davidson.
\newblock {ILR: Where'd My Gadgets Go?}
\newblock In {\em Proceedings of the 33rd IEEE Symposium on Security and Privacy (S\&P'12)}, 2012.

\bibitem{homescu2013profile}
Andrei Homescu, Steven Neisius, Per Larsen, Stefan Brunthaler, and Michael Franz.
\newblock {Profile-guided Automated Software Diversity}.
\newblock In {\em Proceedings of the 2013 IEEE/ACM International Symposium on Code Generation and Optimization (CGO'13)}, 2013.

\bibitem{kil2006address}
Chongkyung Kil, Jinsuk Jun, Christopher Bookholt, Jun Xu, and Peng Ning.
\newblock {Address Space Layout Permutation (ASLP): Towards Fine-Grained Randomization of Commodity Software}.
\newblock In {\em Proceedings of the 22nd Annual Computer Security Applications Conference (ACSAC'06)}, 2006.

\bibitem{pappas2012smashing}
Vasilis Pappas, Michalis Polychronakis, and Angelos~D Keromytis.
\newblock {Smashing the Gadgets: Hindering Return-Oriented Programming Using In-place Code Randomization}.
\newblock In {\em Proceedings of the 33rd IEEE Symposium on Security and Privacy (S\&P'12)}, 2012.

\bibitem{wartell2012binary}
Richard Wartell, Vishwath Mohan, Kevin~W Hamlen, and Zhiqiang Lin.
\newblock {Binary Stirring: Self-randomizing Instruction Addresses of Legacy x86 Binary Code}.
\newblock In {\em Proceedings of the 19th ACM Conference on Computer and Communications Security (CCS'12)}, 2012.

\bibitem{williams2016shuffler}
David Williams-King, Graham Gobieski, Kent Williams-King, James~P Blake, Xinhao Yuan, Patrick Colp, Michelle Zheng, Vasileios~P Kemerlis, Junfeng Yang, and William Aiello.
\newblock {Shuffler: Fast and Deployable Continuous Code Re-Randomization}.
\newblock In {\em Proceedings of the 12th USENIX Symposium on Operating Systems Design and Implementation (OSDI'16)}, 2016.

\bibitem{strackx2009breaking}
Raoul Strackx, Yves Younan, Pieter Philippaerts, Frank Piessens, Sven Lachmund, and Thomas Walter.
\newblock Breaking the memory secrecy assumption.
\newblock In {\em Proceedings of the Second European Workshop on System Security}, pages 1--8, 2009.

\bibitem{JIT-ROP}
Kevin~Z Snow, Fabian Monrose, Lucas Davi, Alexandra Dmitrienko, Christopher Liebchen, and Ahmad-Reza Sadeghi.
\newblock {Just-In-Time Code Reuse: On the Effectiveness of Fine-Grained Address Space Layout Randomization}.
\newblock In {\em Proceedings of the 34th IEEE Symposium on Security and Privacy (S\&P'13)}, 2013.

\bibitem{ahmed2020methodologies}
Salman Ahmed, Ya~Xiao, Kevin~Z Snow, Gang Tan, Fabian Monrose, and Danfeng Yao.
\newblock {Methodologies for Quantifying (Re-)randomization Security and Timing under JIT-ROP}.
\newblock In {\em Proceedings of the 27th ACM Conference on Computer and Communications Security (CCS'20)}, 2020.

\bibitem{XnR}
Michael Backes, Thorsten Holz, Benjamin Kollenda, Philipp Koppe, Stefan N{\"u}rnberger, and Jannik Pewny.
\newblock {You Can Run but You Can't Read: Preventing Disclosure Exploits in Executable Code}.
\newblock In {\em Proceedings of the 21st ACM Conference on Computer and Communications Security (CCS'14)}, 2014.

\bibitem{LR2}
Kjell Braden, Stephen Crane, Lucas Davi, Michael Franz, Per Larsen, Christopher Liebchen, and Ahmad-Reza Sadeghi.
\newblock {Leakage-Resilient Layout Randomization for Mobile Devices}.
\newblock In {\em Proceedings of the 23rd Annual Network and Distributed System Security Symposium (NDSS'16)}, 2016.

\bibitem{Readactor}
Stephen Crane, Christopher Liebchen, Andrei Homescu, Lucas Davi, Per Larsen, Ahmad-Reza Sadeghi, Stefan Brunthaler, and Michael Franz.
\newblock {Readactor: Practical Code Randomization Resilient to Memory Disclosure}.
\newblock In {\em Proceedings of the 36th IEEE Symposium on Security and Privacy (S\&P'15)}, pages 763--780. IEEE, 2015.

\bibitem{uXOM}
Donghyun Kwon, Jangseop Shin, Giyeol Kim, Byoungyoung Lee, Yeongpil Cho, and Yunheung Paek.
\newblock {uXOM: Efficient eXecute-Only Memory on ARM Cortex-M}.
\newblock In {\em Proceedings of the 28th Conference on USENIX Security Symposium (USENIX Security'19)}, 2019.

\bibitem{HideM}
Jason Gionta, William Enck, and Peng Ning.
\newblock {HideM: Protecting the Contents of Userspace Memory in the Face of Disclosure Vulnerabilities}.
\newblock In {\em Proceedings of the 5th ACM Conference on Data and Application Security and Privacy (CODASPY'15)}, 2015.

\bibitem{NORAX}
Yaohui Chen, Dongli Zhang, Ruowen Wang, Rui Qiao, Ahmed~M Azab, Long Lu, Hayawardh Vijayakumar, and Wenbo Shen.
\newblock {NORAX: Enabling Execute-Only Memory for COTS Binaries on AArch64}.
\newblock In {\em Proceedings of the 38th IEEE Symposium on Security and Privacy (S\&P'17)}, 2017.

\bibitem{SECRET}
Mingwei Zhang, Michalis Polychronakis, and R.~Sekar.
\newblock {Protecting COTS Binaries from Disclosure-Guided Code Reuse Attacks}.
\newblock In {\em Proceedings of the 33rd Annual Computer Security Applications Conference (ACSAC '17)}, 2017.

\bibitem{CarrSteve94}
Steve Carr, Kathryn~S. McKinley, and Chau-Wen Tseng.
\newblock {Compiler Optimizations for Improving Data Locality}.
\newblock In {\em Proceedings of the 6th International Conference on Architectural Support for Programming Languages and Operating Systems (ASPLOS VI)}, 1994.

\bibitem{Heisenbyte}
Adrian Tang, Simha Sethumadhavan, and Salvatore Stolfo.
\newblock {Heisenbyte: Thwarting Memory Disclosure Attacks using Destructive Code Reads}.
\newblock In {\em Proceedings of the 22nd ACM Conference on Computer and Communications Security (CCS'15)}, 2015.

\bibitem{NEAR}
Jan Werner, George Baltas, Rob Dallara, Nathan Otterness, Kevin~Z Snow, Fabian Monrose, and Michalis Polychronakis.
\newblock {No-Execute-After-Read: Preventing Code Disclosure in Commodity Software}.
\newblock In {\em Proceedings of the 11th ACM on Asia Conference on Computer and Communications Security (ASIACCS'16)}, 2016.

\bibitem{ZombieGadgets}
Kevin~Z Snow, Roman Rogowski, Jan Werner, Hyungjoon Koo, Fabian Monrose, and Michalis Polychronakis.
\newblock {Return to the Zombie Gadgets: Undermining Destructive Code Reads via Code Inference Attacks}.
\newblock In {\em Proceedings of the 37th IEEE Symposium on Security and Privacy (S\&P'16)}, 2016.

\bibitem{Reassembly23}
Hyungseok Kim, Soomin Kim, Junoh Lee, Kangkook Jee, and Sang~Kil Cha.
\newblock {Reassembly is Hard: A Reflection on Challenges and Strategies}.
\newblock In {\em Proceedings of the 32nd USENIX Conference on Security Symposium (USENIX Security'23)}, 2023.

\bibitem{IntelManual}
Intel.
\newblock {Intel® 64 and IA-32 Architectures Software Developer’s Manual}.
\newblock \url{https://intel.ly/3U3Av2E}, [online].

\bibitem{libmpk}
Soyeon Park, Sangho Lee, Wen Xu, Hyungon Moon, and Taesoo Kim.
\newblock {libmpk: Software Abstraction for Intel Memory Protection Keys (Intel MPK)}.
\newblock In {\em Proceedings of the 2019 USENIX Annual Technical Conference (USENIX ATC'19)}, 2019.

\bibitem{pang2021sok}
{Chengbin Pang and Ruotong Yu and Yaohui Chen and Eric Koskinen and Georgios Portokalidis and Bing Mao and Jun Xu}.
\newblock {SoK: All You Ever Wanted to Know About x86/x64 Binary Disassembly But Were Afraid to Ask}.
\newblock In {\em Proceedings of the 42nd IEEE Symposium on Security and Privacy (S\&P'21)}, 2021.

\bibitem{GroundTruth}
Chengbin Pang, Tiantai Zhang, Ruotong Yu, Bing Mao, and Jun Xu.
\newblock {Ground Truth for Binary Disassembly is Not Easy}.
\newblock In {\em Proceedings of the 31st USENIX Security Symposium (USENIX'22)}, 2022.

\bibitem{lmbench}
Larry McVoy.
\newblock {LMbench - Tools for Performance Analysis}.
\newblock \url{https://lmbench.sourceforge.net/}, [online].

\bibitem{SPEC2006}
Standard Performance~Evaluation Corporation.
\newblock {SPEC CPU 2006}.
\newblock \url{https://www.spec.org/cpu2006/}, [online].

\bibitem{SPEC2017}
Standard Performance~Evaluation Corporation.
\newblock {SPEC CPU® 2017}.
\newblock \url{https://www.spec.org/cpu2017/}, [online].

\bibitem{Roemer12}
Ryan Roemer, Erik Buchanan, Hovav Shacham, and Stefan Savage.
\newblock {Return-Oriented Programming: Systems, Languages, and Applications}.
\newblock {\em ACM Transactions on Information and System Security}, 15(1), March 2012.

\bibitem{multics_xom}
R.~M. Graham.
\newblock {Multics System-Programmers' Manual}.
\newblock \url{https://bit.ly/488lNgQ}, July 1967.

\bibitem{android_pan}
Siguza.
\newblock {PAN}.
\newblock \url{https://blog.siguza.net/PAN/}, [online].

\bibitem{android_xom}
Android.
\newblock {Execute-Only Memory (XOM) for AArch64 Binaries}.
\newblock \url{https://source.android.com/docs/security/test/execute-only-memory}, [online].

\bibitem{kernel_patch}
Dave Hansen.
\newblock {X86, Pkeys: Execute-Only Support}.
\newblock \url{https://lore.kernel.org/linux-mm/20160212210240.CB4BB5CA@viggo.jf.intel.com/}, [online].

\bibitem{intel_kernel_xom}
Rick Edgecombe.
\newblock {Touch But Don’t Look - Running the Kernel in Execute-Only Memory}.
\newblock \url{https://bit.ly/4h2YzfW}, [online].

\bibitem{csw2023}
Theo de~Raadt.
\newblock {Synthetic Memory Protections - An Update on ROP Mitigations}.
\newblock \url{https://www.openbsd.org/papers/csw2023.pdf}, [online].

\bibitem{Meng16}
Xiaozhu Meng and Barton~P. Miller.
\newblock {Binary Code is Not Easy}.
\newblock In {\em Proceedings of the 25th International Symposium on Software Testing and Analysis (ISSTA'16)}, 2016.

\bibitem{Rigger18}
Manuel Rigger, Stefan Marr, Stephen Kell, David Leopoldseder, and Hanspeter M{\"o}ssenb{\"o}ck.
\newblock {An Analysis of x86-64 Inline Assembly in C Programs}.
\newblock In {\em Proceedings of the 14th ACM SIGPLAN/SIGOPS International Conference on Virtual Execution Environments (VEE '18)}, 2018.

\bibitem{KCFI}
Jonathan Corbet.
\newblock {A new LLVM CFI implementation}.
\newblock \url{https://lwn.net/Articles/898040/}, [online].

\bibitem{wartell2011differentiating}
Richard Wartell, Yan Zhou, Kevin~W Hamlen, Murat Kantarcioglu, and Bhavani Thuraisingham.
\newblock {Differentiating Code from Data in x86 Binaries}.
\newblock In {\em Joint European Conference on Machine Learning and Knowledge Discovery in Databases}, 2011.

\bibitem{SEIMI}
Zhe Wang, Chenggang Wu, Mengyao Xie, Yinqian Zhang, Kangjie Lu, Xiaofeng Zhang, Yuanming Lai, Yan Kang, and Min Yang.
\newblock {SEIMI: Efficient and Secure SMAP-Enabled Intra-process Memory Isolation}.
\newblock In {\em Proceedings of the 41st IEEE Symposium on Security and Privacy (S\&P'20)}, 2020.

\bibitem{AModelPopek1978}
Gerald~J. Popek and David~A. Farber.
\newblock A model for verification of data security in operating systems.
\newblock In {\em Communications of the ACM}, 1978.

\bibitem{perfcounter}
XIAO GUANGRONG.
\newblock {Linux Kernel 2.6.31 - 'perf\_counter\_open()' Local Buffer Overflow}.
\newblock \url{https://www.exploit-db.com/exploits/33228}, [online].

\bibitem{NvDriverExp}
Anonymous.
\newblock {Nvidia Linux Driver - Local Privilege Escalation}.
\newblock \url{https://www.exploit-db.com/exploits/20201}, [online].

\bibitem{KernelExploit}
Enrico Perla and Massimiliano Oldani.
\newblock {\em A Guide to Kernel Exploitation: Attacking the Core}.
\newblock Elsevier, 2010.

\bibitem{KHide}
Jason Gionta, William Enck, and Per Larsen.
\newblock {Preventing Kernel Code-Reuse Attacks Through Disclosure Resistant Code Diversification}.
\newblock In {\em Proceedings of 2016 IEEE Conference on Communications and Network Security (CNS'16)}, 2016.

\bibitem{KRX}
Marios Pomonis, Theofilos Petsios, Angelos~D Keromytis, Michalis Polychronakis, and Vasileios~P Kemerlis.
\newblock {kR\^{}X: Comprehensive Kernel Protection Against Just-In-Time Code Reuse}.
\newblock In {\em Proceedings of the Twelfth European Conference on Computer Systems (EuroSys'17)}, 2017.

\bibitem{MPX-dead}
Michael Larabel.
\newblock {Intel MPX Support Will Be Removed From Linux---Memory Protection Extensions Appear Dead}.
\newblock \url{https://www.phoronix.com/news/Intel-MPX-Kernel-Removal-Patch}, 2018.

\bibitem{IskiOS}
Spyridoula Gravani, Mohammad Hedayati, John Criswell, and Michael~L Scott.
\newblock {Fast Intra-kernel Isolation and Security with IskiOS}.
\newblock In {\em Proceedings of the 24th International Symposium on Research in Attacks, Intrusions and Defenses (RAID'21)}, 2021.

\bibitem{PAC}
Qualcomm Technologies.
\newblock {Pointer Authentication on ARMv8.3}.
\newblock \url{https://www.qualcomm.com/content/dam/qcomm-martech/dm-assets/documents/pointer-auth-v7.pdf}, [online].

\bibitem{CET}
Intel.
\newblock {Control Flow Enforcement Technology (CET)}.
\newblock \url{https://www.intel.com/content/dam/develop/external/us/en/documents/catc17-introduction-intel-cet-844137.pdf}, [online].

\bibitem{OutOfControl}
Enes G{\"o}ktas, Elias Athanasopoulos, Herbert Bos, and Georgios Portokalidis.
\newblock Out of control: Overcoming control-flow integrity.
\newblock In {\em Proceedings of the 2014 IEEE Symposium on Security and Privacy (S\&P '14)}, pages 575--589. IEEE, 2014.

\bibitem{ControlFlowBending}
Nicholas Carlini, Antonio Barresi, Mathias Payer, David Wagner, and Thomas~R Gross.
\newblock {Control-Flow Bending: On the Effectiveness of Control-Flow Integrity}.
\newblock In {\em Proceedings of the 24th USENIX Security Symposium (USENIX Security '15)}, pages 161--176, 2015.

\bibitem{StitchingGadgets}
Lucas Davi, Ahmad-Reza Sadeghi, Daniel Lehmann, and Fabian Monrose.
\newblock Stitching the gadgets: On the ineffectiveness of coarse-grained control-flow integrity protection.
\newblock In {\em Proceedings of the 23rd USENIX Security Symposium (USENIX Security '14)}, pages 401--416, 2014.

\bibitem{ControlJujutsu}
Isaac Evans, Fan Long, Ulziibayar Otgonbaatar, Howard Shrobe, Martin Rinard, Hamed Okhravi, and Stelios Sidiroglou-Douskos.
\newblock Control jujutsu: On the weaknesses of fine-grained control flow integrity.
\newblock In {\em Proceedings of the 22nd ACM SIGSAC Conference on Computer and Communications Security (CCS '15)}, pages 901--913, 2015.

\bibitem{StillDangerous}
Nicholas Carlini and David Wagner.
\newblock Rop is still dangerous: Breaking modern defenses.
\newblock In {\em 23rd USENIX Security Symposium (USENIX Security '14)}, pages 385--399, 2014.

\bibitem{SizeDoesMatter}
Enes G{\"o}kta{\c{s}}, Elias Athanasopoulos, Michalis Polychronakis, Herbert Bos, and Georgios Portokalidis.
\newblock Size does matter: Why using gadget-chain length to prevent code-reuse attacks is hard.
\newblock In {\em Proceedings of the 23rd USENIX Security Symposium (USENIX Security '14)}, pages 417--432, 2014.

\bibitem{CFI_JVM}
Sabine Houy and Alexandre Bartel.
\newblock {Lessons Learned and Challenges of Deploying Control Flow Integrity in Complex Software: The Case of OpenJDK’s Java Virtual Machine}.
\newblock In {\em Proceedings of the 2024 IEEE Secure Development Conference (SecDev '24)}, 2024.

\bibitem{mughal2018art}
Arif~Ali Mughal.
\newblock {The Art of Cybersecurity: Defense in Depth Strategy for Robust Protection}.
\newblock {\em International Journal of Intelligent Automation and Computing}, 1(1):1--20, 2018.

\bibitem{DEEPDI}
Sheng Yu, Yu~Qu, Xunchao Hu, and Heng Yin.
\newblock {DeepDi: Learning a Relational Graph Convolutional Network Model on Instructions for Fast and Accurate Disassembly}.
\newblock In {\em Proceedings of the 31st USENIX Security Symposium (USENIX '22)}, 2022.

\bibitem{ProbabilisticDisassembly}
Kenneth Miller, Yonghwi Kwon, Yi~Sun, Zhuo Zhang, Xiangyu Zhang, and Zhiqiang Lin.
\newblock {Probabilistic Disassembly}.
\newblock In {\em Proceedings of the 41st International Conference on Software Engineering (ICSE '19)}, 2019.

\bibitem{DatalogDsiassembly}
Antonio Flores-Montoya and Eric Schulte.
\newblock {Datalog Disassembly}.
\newblock In {\em Proceedings of the 29th USENIX Conference on Security Symposium (USENIX Security'20)}, 2020.

\bibitem{andriesse2016depth}
Dennis Andriesse, Xi~Chen, Victor Van Der~Veen, Asia Slowinska, and Herbert Bos.
\newblock {An In-Depth Analysis of Disassembly on Full-Scale x86/x64 Binaries}.
\newblock In {\em Proceedings of the 25th USENIX security symposium (USENIX security '16)}, pages 583--600, 2016.

\bibitem{bauman2018superset}
Erick Bauman, Zhiqiang Lin, Kevin~W Hamlen, et~al.
\newblock {Superset Disassembly: Statically Rewriting x86 Binaries Without Heuristics}.
\newblock In {\em Proceedings of the 2018 Network and Distributed Systems Security Symposium (NDSS '18)}, 2018.

\bibitem{XDA}
Kexin Pei, Jonas Guan, David Williams-King, Junfeng Yang, and Suman Jana.
\newblock {XDA: Accurate, Robust Disassembly with Transfer Learning}.
\newblock In {\em Proceedings of the 2021 Network and Distributed Systems Security Symposium (NDSS '21)}, 2021.

\bibitem{schwarz2002disassembly}
Benjamin Schwarz, Saumya Debray, and Gregory Andrews.
\newblock {Disassembly of Executable Code Revisited}.
\newblock In {\em Proceedings of 9th Working Conference on Reverse Engineering}, 2002.

\bibitem{Egalito}
David Williams-King, Hidenori Kobayashi, Kent Williams-King, Graham Patterson, Frank Spano, Yu~Jian Wu, Junfeng Yang, and Vasileios~P. Kemerlis.
\newblock {Egalito: Layout-Agnostic Binary Recompilation}.
\newblock In {\em Proceedings of the 25th International Conference on Architectural Support for Programming Languages and Operating Systems (ASPLOS '20)}, 2020.

\bibitem{ROPGadget}
Jonathan Salwan.
\newblock {ROPgadget}.
\newblock \url{https://github.com/JonathanSalwan/ROPgadget}, [online].

\bibitem{schwartz2011q}
Edward~J Schwartz, Thanassis Avgerinos, and David Brumley.
\newblock Q: Exploit hardening made easy.
\newblock In {\em Proceedings of the 20th USENIX Conference on Security Symposium (USENIX Security'11)}, 2011.

\bibitem{Microgadgets}
Andrei Homescu, Michael Stewart, Per Larsen, Stefan Brunthaler, and Michael Franz.
\newblock {Microgadgets: Size Does Matter in Turing-Complete Return-Oriented Programming}.
\newblock In {\em Proceedings of the 6th USENIX Workshop on Offensive Technologies (WOOT '12)}, 2012.

\bibitem{jirop-native}
Salman Ahmed.
\newblock {JITROP-Native}.
\newblock \url{https://github.com/salmanyam/jitrop-native}, [online].

\bibitem{Pitfalls}
R.~Joseph Connor, Tyler McDaniel, Jared~M. Smith, and Max Schuchard.
\newblock {PKU Pitfalls: Attacks on PKU-based Memory Isolation Systems}.
\newblock In {\em Proceedings of the 29th USENIX Security Symposium (USENIX Security '20)}, 2020.

\bibitem{Jenny}
David Schrammel, Samuel Weiser, Richard Sadek, and Stefan Mangard.
\newblock {Jenny: Securing Syscalls for PKU-based Memory Isolation Systems}.
\newblock In {\em Proceedings of the 31st USENIX Security Symposium (USENIX Security '22)}, 2022.

\bibitem{Cerberus}
Alexios Voulimeneas, Jonas Vinck, Ruben Mechelinck, and Stijn Volckaert.
\newblock {You Shall Not (By)pass! Practical, Secure, and Fast PKU-based Sandboxing}.
\newblock In {\em Proceedings of the 17th European Conference on Computer Systems (EuroSys '22)}, 2022.

\bibitem{ERIM}
Anjo Vahldiek-Oberwagner, Eslam Elnikety, Nuno~O. Duarte, Michael Sammler, Peter Druschel, and Deepak Garg.
\newblock {ERIM: Secure, Efficient In-process Isolation with Protection Keys (MPK)}.
\newblock In {\em Proceedings of the 28th Conference on USENIX Security Symposium (USENIX Security'19)}, 2019.

\bibitem{Hodor}
Mohammad Hedayati, Spyridoula Gravani, Ethan Johnson, John Criswell, Michael~L Scott, Kai Shen, and Mike Marty.
\newblock {Hodor: Intra-Process Isolation for High-Throughput Data Plane Libraries}.
\newblock In {\em Proceedings of the 2019 USENIX Annual Technical Conference (USENIX ATC'19)}, 2019.

\bibitem{ShadowStacksMPK}
Nathan Burow, Xinping Zhang, and Mathias Payer.
\newblock {SoK: Shining Light on Shadow Stacks}.
\newblock In {\em Proceedings of the 40th IEEE Symposium on Security and Privacy (S\&P'19)}, 2019.

\bibitem{ShadowStacks}
Thurston~HY Dang, Petros Maniatis, and David Wagner.
\newblock {The Performance Cost of Shadow Stacks and Stack Canaries}.
\newblock In {\em Proceedings of the 10th ACM Symposium on Information, Computer and Communications Security (ASIA CCS'15)}, 2015.

\bibitem{CryptoMPK}
Xuancheng Jin, Xuangan Xiao, Songlin Jia, Wang Gao, Hang Zhang, Dawu Gu, Siqi Ma, Zhiyun Qian, and Juanru Li.
\newblock {Annotating, Tracking, and Protecting Cryptographic Secrets with CryptoMPK}.
\newblock In {\em Proceedings of the 43rd IEEE Symposium on Security and Privacy (S\&P'22)}, 2022.

\bibitem{ApacheBench}
The Apache~Software Foundation.
\newblock {Apache HTTP server benchmarking tool}.
\newblock \url{https://httpd.apache.org/docs/2.4/programs/ab.html}, [online].

\bibitem{sysbench}
Oracle.
\newblock {MySQL Benchmark Tool}.
\newblock \url{https://dev.mysql.com/downloads/benchmarks.html}, [online].

\bibitem{mongo-perf}
MongoDB.
\newblock {Performance Tools for Mongodb}.
\newblock \url{https://github.com/mongodb/mongo-perf}, [online].

\bibitem{redis-benchmark}
Redis Ltd.
\newblock {How fast is Redis?}
\newblock \url{https://redis.io/topics/benchmarks}, [online].

\bibitem{CVE-2013-2028}
The~MITRE Corporation.
\newblock {CVE-2013-2028 Detail}.
\newblock \url{https://www.cve.org/CVERecord?id=CVE-2013-2028}, [online].

\bibitem{usenix24dataonly}
Brian Johannesmeyer, Asia Slowinska, Herbert Bos, and Cristiano Giuffrida.
\newblock {Practical Data-Only Attack Generation}.
\newblock In {\em Proceedings of the 33rd USENIX Conference on Security Symposium (USENIX Security' 24)}, 2024.

\end{thebibliography}
